\documentclass{aa}
\usepackage{graphicx}
\usepackage{txfonts}
\usepackage{natbib}
\usepackage{longtable} 
\begin{document}
\title{Panchromatic study of GRB 060124: from precursor to afterglow}
\author{P. Romano\inst{1},     
	S.~Campana\inst{1},    
	G.~Chincarini\inst{1,2},
	J.~Cummings\inst{3,4},  
	G.~Cusumano\inst{5},    
	S.~T.~Holland\inst{3,6},
	V.~Mangano\inst{5},  
	T.~Mineo\inst{5},   
	K.~L.~Page\inst{7}, 
	V.~Pal'shin\inst{8},
	E.~Rol\inst{7},     
	T.~Sakamoto\inst{3,4}, 
	B.~Zhang\inst{9},    
	R.~Aptekar\inst{8},	
	S.~Barbier\inst{3},  
	S.~Barthelmy\inst{3},
	A.~P.~Beardmore\inst{7},
	P.~Boyd\inst{3},       
	D.~N.~Burrows\inst{10},
	M.~Capalbi\inst{11},   
	E.~E.~Fenimore\inst{12},   
	D.~Frederiks\inst{8},  
	N.~Gehrels\inst{3},  
        P.~Giommi\inst{11},    
	M.~R.~Goad\inst{7},  
	O.~Godet\inst{7},     
	S.~Golenetskii\inst{8},
	D.~Guetta\inst{13},   
	J.~A.~Kennea\inst{10},
	V.~La~Parola\inst{5}, 
	D.~Malesani\inst{14}, 
	F.~Marshall\inst{3}, 
	A.~Moretti\inst{1},  
	J.~A.~Nousek\inst{10},
	P.~T.~O'Brien\inst{7}, 
	J.~P.~Osborne\inst{7}, 
	M.~Perri\inst{11},
	G.~Tagliaferri\inst{1} 
	}
\offprints{P.\ Romano, INAF--Osservatorio Astronomico di Brera, Via E.\ Bianchi 46, I-23807 Merate (LC), Italy; 
		\email{romano@merate.mi.astro.it}}
\institute{INAF--Osservatorio Astronomico di Brera, Via E.\ Bianchi 46, I-23807 Merate (LC), Italy
	\and Universit\`a{} degli Studi di Milano, Bicocca, Piazza delle Scienze 3, I-20126, Milano, Italy
	\and NASA/Goddard Space Flight Center, Greenbelt, MD 20771, USA  
	\and National Research Council, 2101 Constitution Avenue, NW, TJ2114, Washington, DC 20418. 
  	\and INAF--Istituto di Astrofisica Spaziale e Fisica Cosmica Sezione di Palermo,
              Via U.\ La Malfa 153, I-90146 Palermo, Italy   
	\and Universities Space Research Association, 10211 Wincopin Circle, Suite 500, Columbia, MD, 21044-3432, USA
	\and Department of Physics \& Astronomy, University of Leicester, LE1 7RH, UK
	\and Ioffe Physico-Technical Institute, 26 Polytekhnicheskaya, St Petersburg 194021, Russian Federation
  	\and Department of Physics, University of Nevada, Las Vegas, NV 89154-4002, USA
  	\and Department of Astronomy \& Astrophysics, Pennsylvania State University,  
		525 Davey Lab, University Park, PA 16802, USA 
  	\and ASI Science Data Center, via G.\ Galilei, I-00044 Frascati (Roma), Italy 
	\and Los Alamos National Laboratory, MS B244, NM 87545, USA
	\and INAF--Osservatorio Astronomico di Roma, Via di Frascati 33, I-00040 Monteporzio Catone, Italy
	\and International School for Advanced Studies (SISSA-ISAS), Via Beirut 2-4, I-34014 Trieste, Italy 
	}  
\date{Received: 22 February 2006 /  Accepted 15 March 2006 }
\abstract{
We present observations of \object{GRB~060124}, the first event for which 
both the prompt and the afterglow emission could be observed simultaneously 
and in their entirety by the three Swift instruments. 
Indeed, Swift-BAT triggered on a precursor $\sim 570$\,s before the main burst peak,
and this allowed Swift to repoint the narrow field instruments to the burst position 
$\sim 350$\,s {\it before} the main burst occurred. 
GRB~060124 also triggered Konus-Wind, which observed the prompt emission in a 
harder gamma-ray band (up to 2\,MeV).
Thanks to these exceptional circumstances, the temporal and spectral properties 
of the prompt emission can be studied in the optical, X-ray and gamma-ray ranges.
While the X-ray emission (0.2--10~keV) clearly tracks the
gamma-ray burst, the optical component follows a different pattern,
likely indicating a different origin, possibly the onset of external
shocks.
The prompt GRB spectrum shows significant spectral evolution,
with both the peak energy and the spectral index varying. As observed in
several long GRBs, significant lags are measured between the hard- and
low-energy components, showing that this behaviour extends over 3
decades in energy.
The GRB peaks are also much broader at soft energies.
This is related to the temporal evolution of the spectrum, and can be
accounted for by assuming that the electron spectral index softened with
time.
The burst energy ($E_{\rm iso} \sim 5\times 10^{53}$ erg) 
and average peak energy ($E_{\rm p} \sim 300$~keV) 
make GRB\,060124 consistent with the Amati relation.
The X-ray afterglow is characterized by a 
decay which presents a break at $t_{\rm b} \sim 10^5$~s.

}
\authorrunning {P.\ Romano}
\titlerunning {Prompt and afterglow emission of GRB~060124}
\maketitle
%%%%%%%%%%%%%%%%%%%%%%%%%%%%%%%%%%%%%%%%%%%%%%%%%%%%%%%%%%%%%%%%
	\section{Introduction\label{grb060124:introd}}
%%%%%%%%%%%%%%%%%%%%%%%%%%%%%%%%%%%%%%%%%%%%%%%%%%%%%%%%%%%%%%%%

The Swift Gamma-Ray Burst Explorer (\citealt{Gehrels04})
was successfully launched on 2004 Nov.\ 20.
Its payload includes one wide-field instrument, 
the gamma-ray Burst Alert Telescope (BAT; \citealt{BAT}, 15--350 keV energy band),  
and two narrow-field instruments (NFIs), 
the X-Ray Telescope (XRT; \citealt{XRT}, 0.2--10\,keV) 
and the Ultraviolet/Optical Telescope (UVOT; \citealt{UVOT}, 1700--6500\,\AA{}).
The BAT detects the bursts, calculates their positions 
to $\la 3\arcmin$ accuracy and triggers an autonomous 
slew of the observatory to point the two narrow-field instruments,
typically within 100\,s from the burst onset.
The XRT can provide $\la 5\arcsec$ positions, 
while the UVOT further refines the afterglow localization to $\sim 0\farcs5$. 

GRB~060124 is the first event for which the three Swift instruments 
have a clear detection of both the prompt and the afterglow emission. 
Indeed, Swift-BAT triggered on a precursor on 2006-01-24 at 15:54:52 UT, 
$\sim 570$\,s before the main burst peak. 
This allowed Swift to immediately repoint the NFIs
and acquire a pointing towards the burst $\sim 350$\,s {\it before} the burst occurred. 
The burst, which had a highly structured burst profile, comprises 
three major peaks following the precursor and had  one of longest 
total durations (even excluding the precursor) 
recorded by either BATSE or Swift.  

GRB~060124 also triggered Konus-Wind \citep{KonusWind} 
559.4\,s after the BAT trigger \citep{golenetskii2006:gcn4599}. 
The Konus light curve confirmed the presence of both the 
precursor and the three peaks of prompt emission. 
The main peak of GRB 060124 was also bright enough to trigger the FREGATE 
instrument aboard HETE--II (HETE trigger 4012, \citealt{lamb2006:gcn4601})
557.7\,s after the BAT trigger. 

The  prompt emission of GRB~060124 was observed simultaneously by XRT 
with exceptional signal-to-noise (S/N) and was detected by UVOT at 
$V=16.96\pm 0.08$ ($T+183$\,s) and 
$V=16.79\pm 0.04$ ($T+633$\,s). 
This fact makes it an exceptional 
test case to study prompt emission models, since 
this is the very first case that the burst proper 
could be observed with an X-ray CCD with high spatial resolution imaging
down to 0.2\,keV. 
There have been a handful of previous optical detections 
while the burst prompt emission was still active, due to the increasing 
number of robotic followup telescopes.
\object{GRB~990123} was observed by ROTSE \citep{Akerlof1999}, and its light
curve showed a flare which rapidly decayed, uncorrelated with the
gamma-ray emission. This behaviour was generally interpreted as emission
from the external reverse shock \citep{Sari1999:opticalflash,Meszaros1999:990123}.
A reverse shock was also likely observed in \object{GRB~050904} \citep{grb050904opt}. 
A different behaviour was observed for \object{GRB~041219A} 
\citep{Vestrand:041219a,Blake:041219a}.
In this case, the optical emission showed a temporal
variability correlated with the gamma-ray emission, suggesting a common
origin for the two components. 
The prompt afterglows of \object{GRB~050319}
\citep{Quimby2006:050319,Wozniak2005:050319}
 and \object{GRB~050401} \citep{Rykoff2005:050401} showed a decay
matching the late-time one, without the need for extra components. Last,
\object{GRB~050801} \citep{Rykoff2006:050801}  showed a period of
plateaux up to a few minutes after the burst, excluding contribution from
the internal shocks, but possibly arguing for energisation from the
central engine correlated with the X-ray emission.

GRB~060124 was extensively followed from ground-based observatories
with detections starting from one hour after the trigger.
The optical counterpart was identified by the 
Tautenburg telescope \citep{kann2006:gcn4574}. 
Further observations were carried out 
with the ARIES 1.04m \citep{misra2006:gcn4589},
the HCT 2m \citep{bhatt2006:gcn4597}, 
the CrAO 2.6m \citep{rumyantsev2006:gcn4609} and 
the Asiago 1.82m \citep{masetti2006:gcn4587}. 
Limits were given by  ART \citep{torii2006:gcn4596}, 
MASTER \citep{lipunov2006:gcn4572},  UoM 30cm \citep{sonoda2006:gcn4576}, 
TAROT \citep{klotz2006:gcn4581}, and Loiano-1.52m \citep{greco2006:gcn4605}.
An absorption redshift of $z=2.297\pm 0.001$ was determined based on the  
CIV$\lambda\lambda 1548,1550$ doublet 
\citep{mirabal2006:gcn4591,cenko2006:gcn4592,prochaska2006:gcn4593}.

This paper is organized as follows. 
In Sect.~\ref{grb060124:dataredu} we describe our observations and data reduction; 
in Sect.~\ref{grb060124:dataanal} we describe our instrument-based spatial, timing 
and spectral data analysis; 
in Sect.~\ref{grb060124:fulltiming} and \ref{grb060124:fullspectral} we describe our detailed, 
multi-wavelength timing and spectral analysis of the prompt and afterglow emission.
In Sect.~\ref{grb060124:discussion} we discuss our findings. 
Finally, in Sect.~\ref{grb060124:conclusions} we summarise our findings and conclusions. 
Throughout this paper the quoted uncertainties are given at 90\% confidence level 
for one interesting parameter (i.e., $\Delta \chi^2 =2.71$) unless otherwise stated.
Times are referred to the BAT trigger $T_0$, $t=T-T_0$, unless otherwise specified. 
The decay and spectral indices are parameterized as  
$F(\nu,t) \propto t^{-\alpha} \nu^{-\beta}$, 
where $F_{\nu}$ (erg cm$^{-2}$ s$^{-1}$ Hz$^{-1}$) is the 
monochromatic flux as a function 
of time $t$ and frequency $\nu$; we also use $\Gamma = \beta +1$ 
as the photon index, 
$N(E) \propto E^{-\Gamma}$ (ph keV$^{-1}$ cm$^{-2}$ s$^{-1}$). 
We adopt a standard cosmology model with $H_0 = 70$ km s$^{-1}$ Mpc$^{-1}$, 
$\Omega_{\rm M} = 0.3$, $\Omega_\Lambda = 0.7$.

%%%%%%%%%%%%%%%%%%%%%%%%%%%%%%%%%%%%%%%%%%%%%%%%%%%%%%%%%%%%%%%%%%%%%%%%%%%%%%%%%%%%%%%%%%%%%%%%%%%%%%%%
\section{Observations and data reduction\label{grb060124:dataredu}} 
%%%%%%%%%%%%%%%%%%%%%%%%%%%%%%%%%%%%%%%%%%%%%%%%%%%%%%%%%%%%%%%%%%%%%%%%%%%%%%%%%%%%%%%%%%%%%%%%%%%%%%%%
	
%%%%%%%%%%%%%%%%%%%%%%%%%%%%%%%%%%%%%%%%%%%%%%%%%%%%%%%%%%%%%%%%
\subsection{BAT observations\label{grb060124:batobs}} %%%%%%%%%%
%%%%%%%%%%%%%%%%%%%%%%%%%%%%%%%%%%%%%%%%%%%%%%%%%%%%%%%%%%%%%%%%

At 15:54:52 UT, Swift-BAT triggered and localized GRB 060124 
\citep[trigger 178\,750, ][]{holland2006:gcn4570,fenimore2006:gcn4586}.
The on-board location was distributed 21\,s after the trigger, and its position was 
RA(J2000$)=05^{\rm h} 08^{\rm m} 10^{\rm s}$,  
Dec(J2000$)=+69^{\circ} 42^{\prime} 33^{\prime\prime}$,  with an 
uncertainty of 3$^{\prime}$. 
The partial coding was 62\%. 
The spacecraft started to slew 
to the burst position at $t=19$\,s, and settled at 
$t=89$\,s.  
Table~\ref{grb060124:tab_obs} reports the log of the Swift-BAT 
observations that were used for this work. 
Since BAT was triggered by the precursor, the event-by-event data 
which are available from $T_0 - 40~\mathrm{s} < t < T_0 
+ 300~\mathrm{s}$ 
only include the precursor and do not include the main burst.

\begin{figure} %%%%%%%%%%%%%%%%%%%%%%%%%%%%%%%%%%%%%%%%%%%%%%%%%%%%%%%%%%%%
 	 	\resizebox{\hsize}{!}{\includegraphics[angle=0]{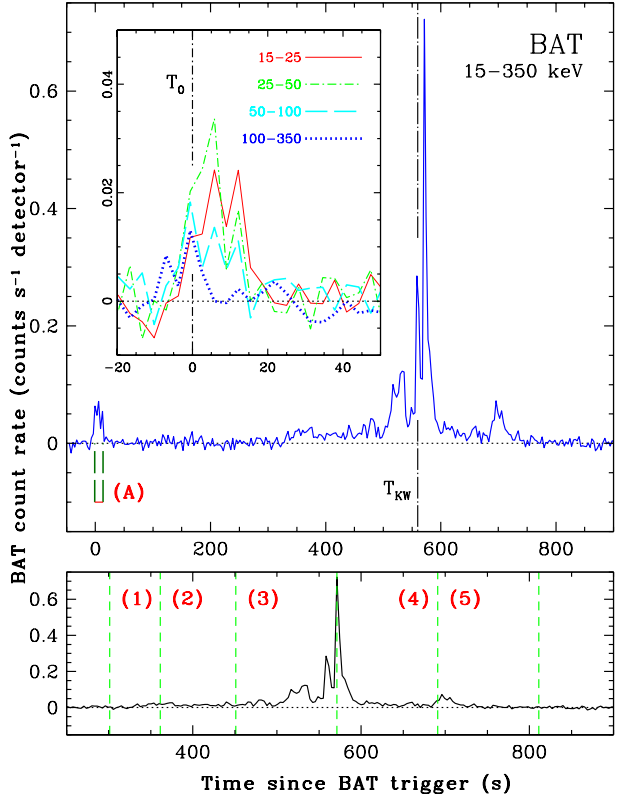}}  
 		\caption{{\bf Top:} Mask-tagged (i.e., background subtracted) BAT light 
		curve of GRB~060124 in the 15--350 keV energy band. 
		The dot-long dash vertical line denotes the Konus trigger time $T_{\rm KW}$.
		The dashed vertical lines denote the boundaries of the interval (A)
		over which we extracted the spectra in Sect.~\ref{grb060124:batanal}. 
		{\bf Inset:} Mask-tagged BAT light curves during the precursor, 
		extracted in the standard 4 energy bands, 
		15--25, 25--50, 50--100, 100--350\,keV,
		at a 3.2\,s time resolution. 
		The dot-short dash vertical line denotes the BAT trigger time $T_0$.
		{\bf Bottom:} The BAT 15--350\,keV light curve around the time of the 
		brightest part of the burst.
		The dashed  vertical lines denote the boundaries of the five intervals 
		over which we extracted the spectra in Sect.~\ref{grb060124:batanal}. 
		}
 		\label{grb060124:batlc}
	\end{figure}

The temporal analysis around the main peak was performed using the 
mask-tagged 4 channel light curves (15--25, 25--50, 50--100, and 
100--350\,keV bands) generated by the flight software.  For the 
spectral analysis, the detector plane histogram (DPH) data 
(80-channel spectral data with time-intervals shown in the 
bottom panel of Fig.~\ref{grb060124:batlc}) were used.  
The BAT mask-weighted light curve (Fig.~\ref{grb060124:batlc}) shows a 
precursor from $t=-3$ to $t=13$\,s,
with a peak count rate of $\sim1700$ counts s$^{-1}$ (15--350 keV) 
at the trigger time,  
then three major peaks: $t=520$--$550$\,s, $t=560$--$580$\,s 
(with the largest flux), and $t=690$--$710$\,s. 
The inset of Fig.~\ref{grb060124:batlc} shows the 16-s precursor.
We note that this is the longest interval ($\approx 500$\,s) 
recorded between 
a precursor and the main GRB event \citep{Lazzati2005:precursors}.

%%%%%%%%%%%%%%%%%%%%%%%%%%%%%%%%%%%%%%%%%%%%%%%%%%%%%%%%%%%%%%%%
\subsection{Konus observations\label{grb060124:kwobs}}
%%%%%%%%%%%%%%%%%%%%%%%%%%%%%%%%%%%%%%%%%%%%%%%%%%%%%%%%%%%%%%%%

	\begin{figure}%%%%%%%%%%%%%%%%%%%%%%%%%%%%%%%%%%%%%%%%%%%%%%%%%%%%%%%%%%%%
  	 	\resizebox{\hsize}{!}{\includegraphics[angle=270]{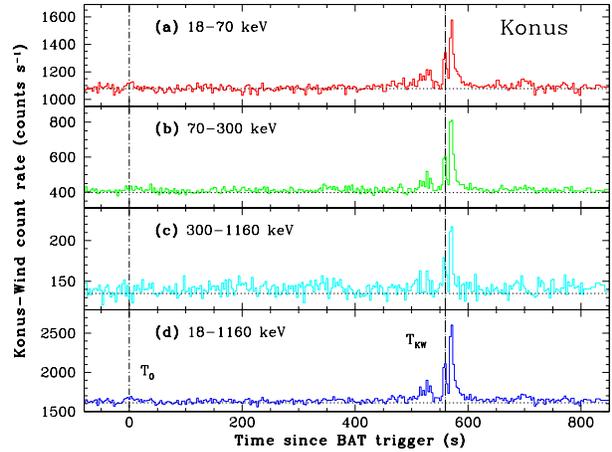}}  
		\caption{Konus light curves of GRB~060124 in the 
		18--70\,keV {\bf (a)}, 70--300\,keV {\bf (b)}, 300-1160\,keV {\bf (c)} and  
		18--1160\,keV {\bf (d)} energy ranges at 2.944\,s time resolution.
		The background levels are marked with dotted horizontal lines. 
		The vertical lines denote the BAT ($T_0$, dot-short dash) and 
		Konus ($T_{\rm KW}$, dot-long dash) trigger times.}
 		\label{grb060124:kwlc}
	\end{figure}

The most intense part of GRB~060124 triggered Konus-Wind 
on 2006-01-24 at $T_{\rm KW}=$16:04:13.9, 559.4\,s after the BAT trigger
(taking into account the $\sim 3$-s propagation delay from Swift to Wind).
It was detected by the S2 detector which observes the North ecliptic
hemisphere; the incident angle was $43\fdg4$. 
The count rates are recorded by Konus in three energy ranges:
G1 (18--70 keV), G2 (70--300 keV), and G3 (300--1160 keV).
The early (trigger) part of the time history is recorded from 
$T_{\rm KW}-$0.512\,s to
$T_{\rm KW}+$229.632\,s with time resolution ranging from 2 to 256\,ms.
The data before $T_{\rm KW}-$0.512\,s are collected in the waiting mode with
2.944\,s time resolution. Time history recorded in the three energy
ranges can be considered as 3-channel spectra.
The full Konus time history of GRB 060124 is reported in Fig.~\ref{grb060124:kwlc},
which shows a weak (4-$\sigma$ and 2.5-$\sigma$ in the 18--70\,keV and 70--300\,keV 
energy bands, respectively) 
precursor corresponding to the Swift-BAT trigger time, 
no statistically significant emission in any Konus energy band 
from $t=20$ to $340$\,s, 
resuming emission at $t=340$\,s, 
the main double-peaked pulse in the interval $t=550$--$590$\,s, 
and weaker pulses up to $t=800$\,s.

%%%%%%%%%%%%%%%%%%%%%%%%%%%%%%%%%%%%%%%%%%%%%%%%%%%%%%%%%%%%%%%%
\subsection{XRT observations\label{grb060124:xrtobs}}
%%%%%%%%%%%%%%%%%%%%%%%%%%%%%%%%%%%%%%%%%%%%%%%%%%%%%%%%%%%%%%%%

The XRT observations of GRB 060124 started on 
2006-01-24 15:56:36 UT, only 104\,s after the BAT trigger, 
 and lasted 
until 2006-02-23 23:32:57 UT. 
Table~\ref{grb060124:tab_obs} reports the log of the Swift-XRT 
observations that were used for this work,  
which include a total net exposure time of $\sim 867$\,s in 
Windowed Timing (WT) mode and $\sim 247$\,ks
in Photon Counting (PC) mode, spread over a $\sim $31\,d baseline. 
The monitoring is organized in 24 sequences. 
The first (000) was performed as an automated target (AT) and 
with XRT in auto state. 
Therefore, in sequence 000 the XRT took an initial 2.5\,s image (IM at $t=104$\,s) 
followed by WT frames (at $t=112$\,s)
for the remainder of the first snapshot (continuous pointing at the target). 
During most of the first sequence, the count rate was sufficiently high for XRT 
to remain in WT mode, the exception being 15 s recorded in PC mode 
\citep[for a description of read-out modes, see][]{Hill04}.
The latter were used to provide a first localization of the afterglow 
\citep{mangano2006:gcn4578}. 
The following observations were performed in both PC and WT modes, but the WT data 
have a very low S/N and were therefore not used here.

%%%%%%%%%%%%%%%%%%%%%%%%%%%%%%%%%%%%%%%%%%%%%%%%%%%%%%%%%%%%%%%%
\subsection{UVOT observations\label{grb060124:uvotobs}}
%%%%%%%%%%%%%%%%%%%%%%%%%%%%%%%%%%%%%%%%%%%%%%%%%%%%%%%%%%%%%%%%

\begin{figure}%%%%%%%%%%%%%%%%%%%%%%%%%%%%%%%%%%%%%%%%%%%%%%%%%%%%%%%%%%%%%%%%%%% 
   	 	\resizebox{\hsize}{!}{\includegraphics[angle=270]{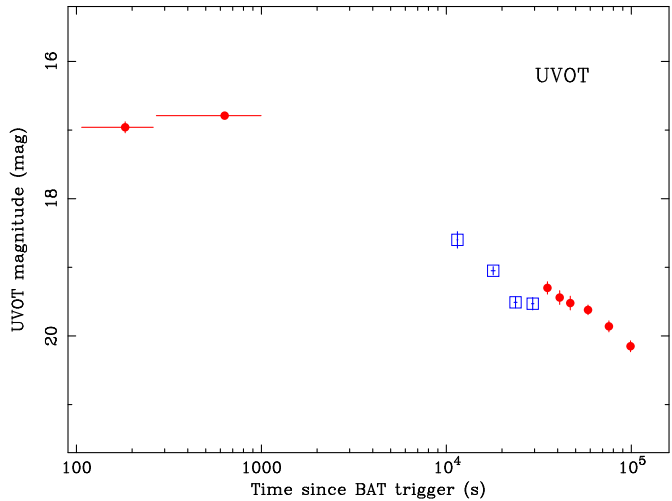}}  
		\caption{UVOT light curves. The $V$ and $B$ magnitudes are shown 
		as (red) filled circles and (blue) empty squares, respectively.
		}
 		\label{grb060124:uvotlc}
	\end{figure}

Table~\ref{grb060124:tab_obs} reports the log of the Swift-UVOT 
observations that were used for this work.
The first UVOT observation was taken in IMAGE mode, because the 
burst interrupted a Target-of-Opportunity (ToO) pointing which
had required this observing mode.
Therefore, the UVOT did not execute the standard AT observing sequence, 
which uses EVENT mode to provide high time resolution.  
By contrast, the IMAGE mode data were collected in long integrations.
After the first observation, the normal follow-up mode was used. 
Figure~\ref{grb060124:uvotlc} shows the observations for 
which a detection was obtained (the upper limits are not shown because 
they do not usefully constrain the behaviour).

%%%%%%%%%%%%%%%%%%%%%%%%%%%%%%%%%%%%%%%%%%%%%%%%%%%%%%%%%%%%%%%%%%%%%%%%%%%%%%%%%%%%%%%%%%%%%%%%%%%%%%%%%%%%%%%%%%%%
\section{Data analysis\label{grb060124:dataanal}} %%%%%%%%%%%%%%
%%%%%%%%%%%%%%%%%%%%%%%%%%%%%%%%%%%%%%%%%%%%%%%%%%%%%%%%%%%%%%%%

	%%%%%%%%%%%%%%%%%%%%%%%%%%%%%%%%%%%%%%%%%%%%%%%%%%%%%%%%%%%%%%%%
	\subsection{BAT and Konus analysis\label{grb060124:batanal}} %%%%%%%%%
	%%%%%%%%%%%%%%%%%%%%%%%%%%%%%%%%%%%%%%%%%%%%%%%%%%%%%%%%%%%%%%%%

\begin{figure}%%%%%%%%%%%%%%%%%%%%%%%%%%%%%%%%%%%%%%%%%%%%%%%%%%%%%%%%%%%%%%%%%%% 
  	 	\resizebox{\hsize}{!}{\includegraphics[angle=0]{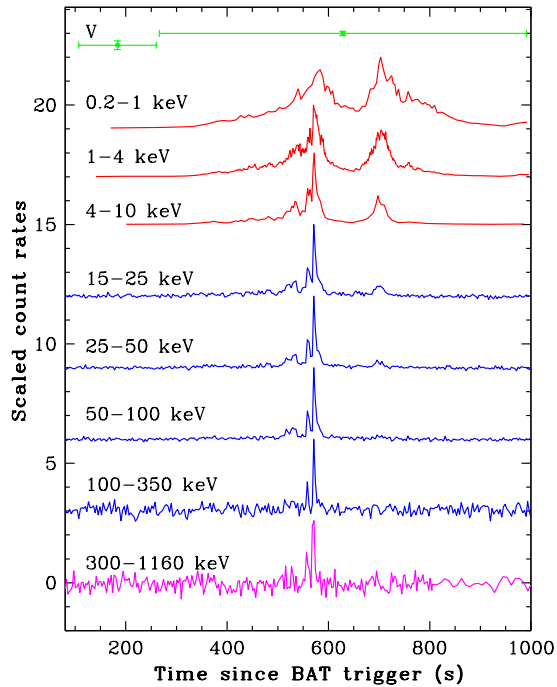}}  
		\caption{UVOT ($V$), XRT, BAT and Konus light curves. 
		The XRT light curves are corrected for pile-up. 
		The count rates have been normalized to the peak of each light curve
		and offset vertically for clarity. 
		}
 		\label{grb060124:scaled}
	\end{figure}

The BAT data were analysed using the standard BAT analysis 
software distributed within FTOOLS v6.0.4. 
The BAT ground-calculated position, based on data from $t=-40$ to $+19$\,s, 
is RA(J2000$)=05^{\rm h} 08^{\rm m} 30\fs6$,  
Dec(J2000$)=+69^{\circ} 43^{\prime} 27\farcs9$, 
with an uncertainty of $1\farcm6$ 
(radius, statistical and systematic errors). 
Mask-tagged BAT light curves were created in the standard 4 energy bands, 
15--25, 25--50, 50--100, 100--350 keV (Fig.~\ref{grb060124:scaled})
and in the total band 15--350 keV (Fig.~\ref{grb060124:batlc})
at 3.2-s time resolution. 
The BAT spectra were extracted over the interval 
which includes the precursor (A in  Fig.~\ref{grb060124:batlc}, 
$t=-1.5$ to $13.5$\,s), and in the other 5 time intervals.
Response matrices were generated with the task {\tt batdrmgen} 
using the latest spectral redistribution matrices. 
For our spectral fitting (XSPEC v11.3.2) 
we considered the 14--150\,keV energy range and applied an energy-dependent 
systematic error 
vector\footnote{http://heasarc.gsfc.nasa.gov/docs/swift/analysis/bat\_digest.html}.

The Konus data were processed using standard Konus analysis tools. 
From $T_{\rm KW}$ to $T_{\rm KW}+491.8$\,s, 64 spectra in 101 channels were
accumulated on timescales varying from 64\,ms near the trigger (first four
spectra) to 8.2~s afterwards.
We also derived the Konus 3-channel spectra and the corresponding response
for several time intervals prior to $T_{\rm KW}$, as well as for the total
burst, for the joint fitting with BAT and XRT.

$T_{90}$ and $T_{50}$ (the timescales over which 90\% and 50\% of the 
burst fluence is measured) were calculated from the BAT light curves with the task
{\tt battblocks}, using data binned to 1.6\,s resolution in the time range 
$t=19.2$--1194.4\,s (thus excluding the precursor) 
and are reported in Table~\ref{grb060124:tabt90s}. 
These were also calculated for Konus data with Konus-specific tasks.
We note, that the Konus $T_{90}$ and $T_{50}$ in the G3 energy band 
are affected by low S/N effects and therefore are probably 
underestimated. 

Although the individual peak durations are 
comparable to the duration of other long GRBs, the total duration 
(excluding the precursor) is among  
the longest recorded by either BATSE \citep{Paciesa1999:batsecatalog} or Swift. 
In the 15--150 keV band the fluences of the precursor emission, 
the main emission and the following peaks are
$(4.7 \pm 0.5) \times 10^{-7}$,  $(1.25\pm 0.03) \times 10^{-5}$ 
and  $(1.05 \pm 0.11) \times 10^{-6}$ erg cm$^{-2}$, respectively.

	%%%%%%%%%%%%%%%%%%%%%%%%%%%%%%%%%%%%%%%%%%%%%%%%%%%%%%%%%%%%%%%%
	\subsection{XRT analysis\label{grb060124:xrtanal}} 
	%%%%%%%%%%%%%%%%%%%%%%%%%%%%%%%%%%%%%%%%%%%%%%%%%%%%%%%%%%%%%%%%

The XRT data were first processed by the Swift Data Center at NASA/GSFC into
Level 1 products (event lists). Then 
they were further processed with the XRTDAS (v1.7.1) software package, written by the
ASI Science Data Center (ASDC) and distributed within
FTOOLS to produce the final cleaned event lists. 
In particular, we ran the task {\tt xrtpipeline} (v0.9.9) applying calibration and 
standard filtering and screening criteria. 
An on-board event threshold of $\sim$0.2\,keV 
was applied to the central pixel of each
event, which has been proven to reduce most of the background due to either
the bright Earth limb or the CCD dark current (which depends on the CCD temperature). 
For our analysis we selected XRT grades 0--12 and 0--2 for PC and WT data, 
respectively (according to Swift nomenclature; \citealt{XRT}). 

The X-ray counterpart was detected at the position 
RA(J2000$)=05^{\rm h} 08^{\rm m} 26\fs03$,  
Dec(J2000$)=+69^{\circ} 44^{\prime} 26\farcs7$, 
with an estimated uncertainty of $3\farcs5$.
This position was determined using the {\tt xrtcentroid} task (v0.2.7) 
on the sequence 002, which is not affected by pile-up, and it takes 
into account the correction for the misalignment
between the telescope and the satellite optical axis.
It is $63\arcsec$ from the refined BAT position 
(Sect.~\ref{grb060124:batanal}).

\begin{figure}%%%%%%%%%%%%%%%%%%%%%%%%%%%%%%%%%%%%%%%%%%%%%%%%%%%%%%%%%%%%%%%%%%% FIGUR5
	%\sidecaption
		\resizebox{\hsize}{!}{\includegraphics[angle=270]{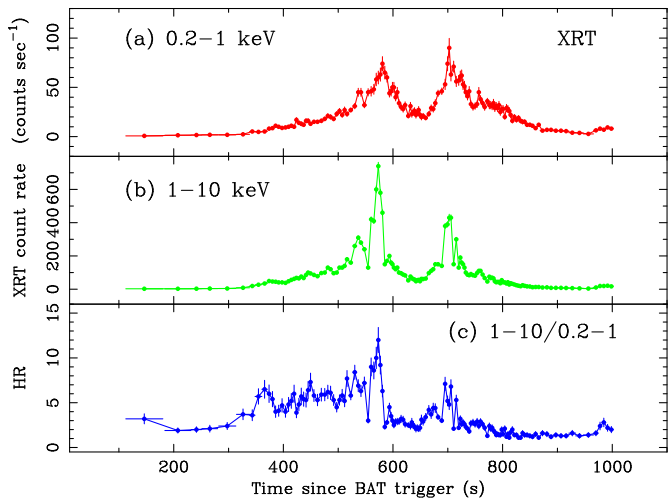}}
		\caption{X-ray WT mode background-subtracted light curves of GRB 060124.
		{\bf (a)}: Soft band ({\bf S}, 0.2--1 keV).
		{\bf (b)}: Hard band ({\bf H}, 1--10 keV). 
		{\bf (c)}: Ratio of hard to soft count rates. 
		}
 		\label{grb060124:xrtwtlc}
	%\end{figure*}
	\end{figure}

During sequence 000 the count rate of the burst was high enough 
to cause pile-up in the WT mode data. 
Therefore, to account for this effect, the WT data were 
extracted in a rectangular 40$\times$20-pixel region 
with a 4$\times$20-pixel region excluded from its centre.
The size of the exclusion region was determined following the 
procedure illustrated in the Appendix.
%, 
To account for the background, WT events were also extracted within a 
rectangular box (40$\times$20 pixels) far from background sources. 
WT events were extracted in three energy bands, 
0.2--10\,keV (total), 0.2--1\,keV (soft, S) and 1--10\,keV (hard, H). 
We dynamically subtracted their respective backgrounds.  
Figure~\ref{grb060124:xrtwtlc} shows the soft and hard 
(background-subtracted and corrected for Point-Spread Function, 
PSF, losses) light curves, as well as the ratio H$/$S, with the 
BAT trigger as the origin of time. 
We also created WT light curves in the 1--4 and 4--10\,keV bands. 
These are shown in Fig.~\ref{grb060124:scaled}. 
$T_{90}$ and $T_{50}$ were calculated from the XRT/WT light curves and 
are reported in Table~\ref{grb060124:tabt90s}.

During most of sequence 001 the PC mode data were piled-up, as well. 
Therefore, we extracted the source events in an annulus with a 35-pixel 
outer radius ($\sim83$\arcsec) and a 3-pixel inner radius. 
These values were derived by comparing the observed and nominal PSF
\citep{vaughan2006:050315}. 
Given the decreasing brightness of the source, for the PC data 
collected in sequence 002 and the following ones
we extracted the spectra in circular regions of decreasing radius 
(35 to 10 pixels) to maximise the S/N. 
PC background data were also extracted in a source-free  
circular region (radius 42 pixels).

Ancillary response files were generated with the task {\tt xrtmkarf} 
within FTOOLS, and account for different extraction regions and 
PSF corrections. 
We used the latest spectral redistribution matrices in the Calibration Database
(CALDB 2.3) maintained by HEASARC.

	%\begin{figure*}%%%%%%%%%%%%%%%%%%%%%%%%%%%%%%%%%%%%%%%%%%%%%%%%%%%%%%%%%%%%%%%%%%% FIGUR5
%

	%%%%%%%%%%%%%%%%%%%%%%%%%%%%%%%%%%%%%%%%%%%%%%%%%%%%%%%%%%%%%%%%
	\subsection{UVOT analysis\label{grb060124:uvotanal}} 
	%%%%%%%%%%%%%%%%%%%%%%%%%%%%%%%%%%%%%%%%%%%%%%%%%%%%%%%%%%%%%%%%

The UVOT located the afterglow of GRB~060124 at 
RA(J2000$)=05^{\rm h} 08^{\rm m} 25\fs859$,  
Dec(J2000$)=+69^{\circ} 44^{\prime} 27\farcs41$, 
with an internal accuracy of $0\farcs01$ and an estimated absolute
astrometric uncertainty of  $0\farcs5$. This is 
$1\farcs1$ from the XRT coordinates (Sect.~\ref{grb060124:xrtanal}). 
To avoid contamination from a $V = 15.4$ source located $7\arcsec$ 
from GRB~060124, we performed aperture photometry using a circular
aperture with a radius of $2\arcsec$ centred on the optical afterglow.
A sky annulus of width $5\arcsec$ and inner radius $17\farcs5$ was
used since this annulus encloses a large sample of sky pixels, and it avoids
the bright sources near the optical afterglow.  Aperture corrections
were performed to convert the $2\arcsec$ photometry aperture into the
standard aperture radii used to define the photometric zero points.
Six isolated stars were used to compute aperture corrections for each
exposure.  
The RMS scatter in the aperture corrections is typically 0.02 mag.

The instrumental magnitudes were transformed to Vega magnitudes using the
photometric zero points in the Swift-UVOT Calibration Database
(CALDB 2.3).  Colour terms have not been applied,
but preliminary calibrations suggest that they are negligible. 
The adopted photometric zero points were $ZP_{V}= 17.88 \pm 0.09$, 
$ZP_{B} = 19.16 \pm 0.12$, $ZP_{UVM2} = 17.29 \pm 0.23$, and  
$ZP_{UVW2} = 17.77 \pm 0.02$.
Table~\ref{grb060124:uvot_mag} reports the photometry.
The UVOT fluxes reported in Table~\ref{grb060124:uvot_mag} 
are monochromatic fluxes computed at the
central wavelength of the appropriate filter.

Figure~\ref{grb060124:uvotlc} shows the Swift-UVOT $V$ and $B$ light 
curves for all observations where a detection of the afterglow was achieved. 
The first two $V$-band points, at 183 and 633\,s, differ
from each other by 0.2\,mag, which corresponds to a 2.3-$\sigma$
variation.  
The UVOT light curve shows a remarkable amount of flickering
(also see Sect.~\ref{grb060124:fulltiming:ag}). 
We note that since these observations were performed in IMAGE mode,
finer time resolution is not available for these data. Therefore, 
the actual, possibly variable behaviour of the optical light curve 
during the prompt phase remains unknown. 
We can estimate that if the prompt emission in the optical 
had occurred only in a time interval comparable to the one of the X-ray 
(and with no detectable emission outside, i.e. the effective exposure 
time is shorter), 
then the UVOT measurement would have been 1.7 mag brighter. 
Even with the (ad-hoc) estimate for this shorter interval the
optical flux would not be larger by the factor
$\sim 100$ by which the X- and gamma-ray
flux increase in the same amount of time. 
In order for the optical emission to 
fully track the hard energy band behaviour, a timescale of emission 
much shorter ($\la 10$\,s) than that observed in the X-ray would have been required.

	%%%%%%%%%%%%%%%%%%%%%%%%%%%%%%%%%%%%%%%%%%%%%%%%%%%%%%%%%%%%%%%%
	\section{Multi-wavelength timing\label{grb060124:fulltiming}} 
	%%%%%%%%%%%%%%%%%%%%%%%%%%%%%%%%%%%%%%%%%%%%%%%%%%%%%%%%%%%%%%%%

	%%%%%%%%%%%%%%%%%%%%%%%%%%%%%%%%%%%%%%%%%%%%%%%%%%%%%%%%%%%%%%%%
	\subsection{The prompt phase\label{grb060124:fulltiming:prompt}}
	%%%%%%%%%%%%%%%%%%%%%%%%%%%%%%%%%%%%%%%%%%%%%%%%%%%%%%%%%%%%%%%%

The prompt phase of GRB~060124 was observed by  UVOT, XRT, BAT and Konus, 
as shown in Fig.~\ref{grb060124:scaled}, which illustrates the prompt 
emission in different energy bands.  
A few observational features can be noticed. 

First, the relative importance of the three main peaks 
(as can be represented by the count rate ratios) varies through the prompt phase, 
in the sense that the third peak flux increases, relative to the second, as the 
energy band becomes softer. For example, the third peak in 
the 0.1--1\,keV light curve is actually stronger than the second one, 
which indicates that strong spectral evolution is taking place
(also, see Fig.~\ref{grb060124:xrtwtlc}c).
This has often been observed before \citep[e.g., ][]{Ford1995:softhard}.

\begin{figure}%%%%%%%%%%%%%%%%%%%%%%%%%%%%%%%%%%%%%%%%%%%%%%%%%%%%%%%%%%%%%%%%%%% FIGURE4
		\resizebox{\hsize}{!}{\includegraphics[angle=270]{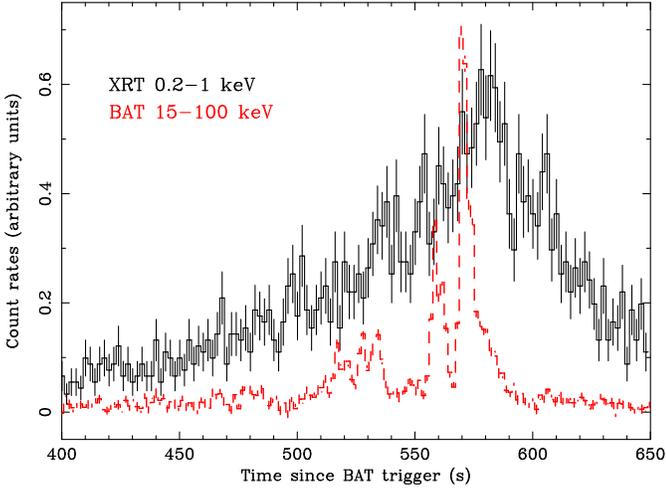}}
 		\caption{Shift of the emission peak with energy band. The XRT 0.2--1\,keV light curve is
		compared with the BAT 15--100\,keV light curve.	The lag between the peaks is $\sim 10$\,s.	
		}
 		\label{grb060124:xrt_02_1kev_bat}
	\end{figure}

Second, the peak times of the emission shift with the energy band, 
as is shown in detail in Fig.~\ref{grb060124:xrt_02_1kev_bat}, 
which shows the 0.2--1\,keV XRT light curve compared with the BAT 15--100\,keV one.
In this extreme case, the lag between the peaks is $\sim 10$\,s. 

\begin{figure}%%%%%%%%%%%%%%%%%%%%%%%%%%%%%%%%%%%%%%%%%%%%%%%%%%%%%%%%%%%%%%%%%%% 
		\resizebox{\hsize}{!}{\includegraphics[angle=270]{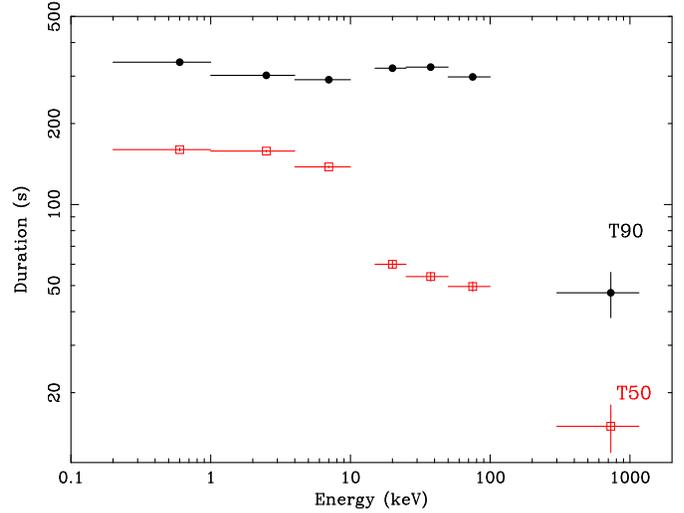}}
 		\caption{Variation of $T_{90}$ (black filled circles) and $T_{50}$ 
		(red empty squares) with energy. 
		The $T_{90}$  and $T_{50}$ error bars correspond to the light curve binning
		(with the exception of the points at $\sim 700$ keV).
		}
 		\label{grb060124:t90}
	\end{figure}

The emission starts off simultaneously in all bands 
(albeit with the shift described above), although it has a considerably 
lower S/N in the BAT and Konus bands 
(see Fig.~\ref{grb060124:batlc} and Fig.~\ref{grb060124:kwlc}
where the total light curves are shown). 
Furthermore, the emission becomes less spiky 
as the energy of the band decreases. 
This is more quantitatively described by the XRT and BAT $T_{90}$ and $T_{50}$ values
which are reported in Table~\ref{grb060124:tabt90s} and illustrated in 
Fig.~\ref{grb060124:t90}. 
While the $T_{90}$s are within 12\% (fractional), the $T_{50}$s 
vary by as much as 42\%. 
This means that the peaks are much broader in the softer bands. 

The X-ray peaks can be formally modelled with 12 Gaussians.
When modelled with power laws, the prompt emission peaks show very 
high rise slopes, ranging from 5 to 12 if the time is referred 
to the BAT trigger, of the order of 1--2 if the initial time is 
set to the first point of the flare.
We note that there is evidence of continued flaring after the end of 
the first snapshot (see Figs.~\ref{grb060124:xrtwtlc} and 
~\ref{grb060124:uvotlc2}), and our $T_{90}$s and $T_{50}$s may 
therefore be underestimated.

	%%%%%%%%%%%%%%%%%%%%%%%%%%%%%%%%%%%%%%%%%%%%%%%%%%%%%%%%%%%%%%%%
	\subsection{The afterglow phase\label{grb060124:fulltiming:ag}}
	%%%%%%%%%%%%%%%%%%%%%%%%%%%%%%%%%%%%%%%%%%%%%%%%%%%%%%%%%%%%%%%%

The afterglow phase was observed by XRT and UVOT.
Since the XRT data are characterized by detection over a longer time 
and with a finer sampling,
a joint fit is strongly biased in favour of the X-ray data. 
Therefore, we fit the XRT and UVOT data separately. 

The XRT/PC portion of the light curve ($t> 10^4$\,s) 
was first fit with a simple power-law model using the BAT trigger as origin of time
(we perform a fit with a different origin, $T_{\rm KW}$, below). 
The fit yields a decay index $\alpha=1.36\pm0.02$ and $\chi^2_{\rm red}=2.43$ 
(94 degrees of freedom, d.o.f.). 
A fit with a broken power law $F(t) = K t^{-\alpha_1}$ for $t<t_{\rm b}$ and  
$F(t) = K\,t_{\rm b}^{-\alpha_1} \, (t/t_{\rm b})^{-\alpha_2}$ for $t>t_{\rm b}$,  
where $t_{\rm b}$ is the time of the break, yields 
$\alpha_1=1.21\pm0.04$ and $\alpha_2=1.58\pm0.06$, 
and a break at $t_{\rm b}=(1.05_{-0.14}^{+0.17})\times 10^{5}$\,s
($\chi^2_{\rm red}=1.87$, 92 d.o.f.). 
This is a significant improvement over the simple power law 
(null hypothesis probability $1.9\times 10^{-6}$).

We note that the XRT/PC light curve is affected by flickering,
so that the residuals to the fits are dominated by rapid variations.
Their intensity decreases as the time increases, a behaviour 
which has been observed before 
and which is due to both the logarithmic binning in our light curves and 
to the power-law decrease of the source counts.
We tested the effect of 
flickering by removing the points in the light curve which were 
at least 2 $\sigma$ above the model being tested for.
Although the normalization of both the power-law and the 
broken power-law fits decrease, the slopes are consistent with 
the previous values within the uncertainties, and $\chi^2_{\rm red}$ 
drops down to $\sim1.2$.

It is also interesting to fit the light curves considering an initial time 
different from the BAT trigger time, i.e. the beginning of the main peak.
To this end, we performed a fit of the PC data, 
using a broken power-law model with an initial time fixed to the Konus 
trigger time $T_{\rm KW}=559.4$\,s. We obtain
$\alpha_1=1.20_{-0.05}^{+0.03}$,
$\alpha_2=1.57_{-0.06}^{+0.07}$, 
and break at $t_{\rm b}=(1.05\pm0.15)\times 10^{5}$\,s
($\chi^2_{\rm red}= 1.88$, 92 d.o.f.),
which is consistent with what we found for the fits which adopted 
the BAT trigger as the starting time.

\begin{figure}%%%%%%%%%%%%%%%%%%%%%%%%%%%%%%%%%%%%%%%%%%%%%%%%%%%%%
 	 	\resizebox{\hsize}{!}{\includegraphics[angle=0]{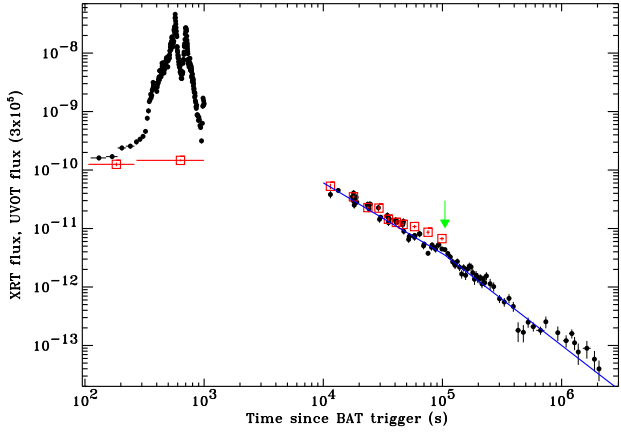}}  
		\caption{XRT (filled circles) light curve in units of 
		erg s$^{-1}$ cm$^{-2}$ and UVOT (empty squares) flux density 
		(erg s$^{-1}$ cm$^{-2}$ \AA$^{-1}$) scaled to match the X-ray. 
		The solid line is the best fit model to the XRT 
		afterglow data. The arrow marks the break position. 
		}
 		\label{grb060124:uvotlc2}
	\end{figure}

The late-time UVOT light curve ($t \ge 11\,478$\,s)  
is characterized by $\alpha_V =0.72 \pm 0.15$ in the $V$ band 
and $\alpha_B =1.03  \pm 0.23 $ in the $B$ band.  
A simultaneous fit of $V$ and $B$ band data with the 
constraint of a single slope yields $\alpha_{BV} = 0.82\pm 0.06$ (68\% error, 
$\chi^2_{\rm red}= 1.13$, 7 d.o.f.). 
We infer an observed value of the colour $B-V=0.47\pm0.08$ 
and an optical spectral index $\beta_{\rm opt}=0.5\pm0.3$ 
(the latter was derived using an extinction correction of $A_{V}=0.447$\,mag).
Figure~\ref{grb060124:uvotlc2} shows the homogenized Swift-UVOT light curve 
(the wavelength dependency of the UVOT bands was taken into account)
superimposed (by scaling) on the XRT flux light curve.

 	%%%%%%%%%%%%%%%%%%%%%%%%%%%%%%%%%%%%%%%%%%%%%%%%%%%%%%%%%%%%%%%%
	\section{Broad-band spectroscopy: XRT, BAT and Konus\label{grb060124:fullspectral}} 
	%%%%%%%%%%%%%%%%%%%%%%%%%%%%%%%%%%%%%%%%%%%%%%%%%%%%%%%%%%%%%%%%

	%%%%%%%%%%%%%%%%%%%%%%%%%%%%%%%%%%%%%%%%%%%%%%%%%%%%%%%%%%%%%%%%
	\subsection{The prompt phase\label{grb060124:multispecfits}} 
	%%%%%%%%%%%%%%%%%%%%%%%%%%%%%%%%%%%%%%%%%%%%%%%%%%%%%%%%%%%%%%%%

As a way to characterize the fundamental burst parameters, we 
used the full Swift and Konus energy ranges. 
To this end we considered the 6 time intervals 
described in Sect.~\ref{grb060124:batanal} and extracted spectra 
from BAT in interval A, 
from XRT and BAT data in intervals 1 and 2, 
and XRT, BAT and Konus in intervals 3--5. 
We fit all the spectra extracted in the same time interval simultaneously. 
We considered an absorbed power law model, with two absorption components, 
the Galactic one $N_{\rm H}^{\rm G}=9.15 \times 10^{20}$ cm$^{-2}$, 
and one at the GRB redshift ($z=2.297$, \citealt{cenko2006:gcn4592}). 
$N_{\rm H}^{\rm G}$ was free to vary between $7.65\times 10^{20}$
and $9.90 \times 10^{20}$ cm$^{-2}$. 
The lower limit was derived from the Galactic extinction estimate 
along the GRB line of sight 
$A_V=0.45$ mag \citep{Schlegelea98}, with a conversion into Hydrogen 
column $N_{\rm H} = 1.7 \times 10^{21} A_V$ cm$^{-2}$ \citep{avnh}.
The second limit is the upper limit of neutral Hydrogen maps 
\citep{DL90}, calculated with the {\tt nh} tool within FTOOLS.   
We note that the Leiden/Argentine/Bonn (LAB) Survey of Galactic HI \citep{LABS}
provides $N_{\rm H}^{\rm G}=9.25 \times 10^{20}$ 
cm$^{-2}$, which is within the considered range. 
We considered different emission models, ranging from a 
simple power-law, 
       $N(E) \propto E^{-\Gamma}$, 
a cutoff power-law,  
       $N(E) \propto E^{-\Gamma} \exp(-E/E_0)$
and the {\tt GRBM} model \citep{Band93},
       $N(E) \propto E^{-\Gamma_1}\,\exp(-E/E_0)$ for $E \le (\Gamma_2-\Gamma_1)E_0$, 
       $N(E) \propto E^{-\Gamma_2}$               for $E > (\Gamma_2-\Gamma_1)E_0$,
where $E_0$ is the cutoff energy, and 
$\Gamma_1$ and $\Gamma_2$ are the low- and high-energy photon indices.
The fits were performed in the 0.5--10, 14--150, 
20--2\,000\,keV energy ranges for XRT/WT, BAT and Konus, respectively.

	\begin{figure}%%%%%%%%%%%%%%%%%%%%%%%%%%%%%%%%%%%%%%%%%%%%%%%%%%%%%%%%%%%%%%
		\resizebox{\hsize}{!}{\includegraphics[angle=270]{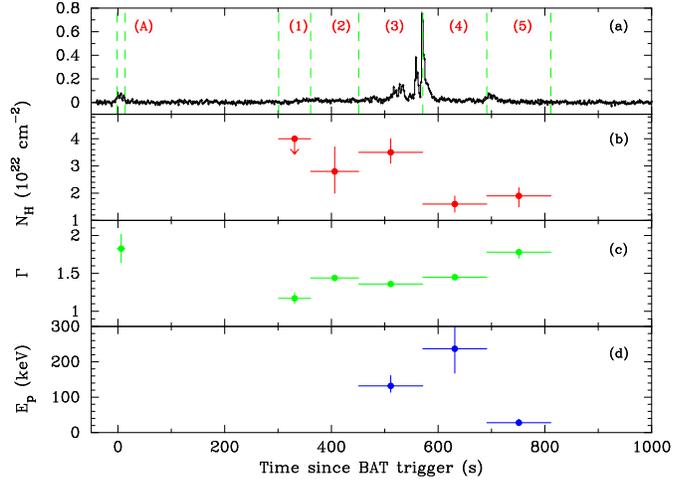}}
  		\caption{Multi-wavelength time-resolved spectral analysis. 
			The models and fit results are reported in 
			Table~\ref{grb060124:multispectab}. 
		{\bf (a)}: BAT  15--350\,keV light curve with the 6 time intervals chosen 
				for the analysis marked; 
		{\bf (b)}: $N_{\rm H}^{z}$ in units of $10^{22}$ cm$^{-2}$; 
		{\bf (c)}: photon index $\Gamma$;
		{\bf (d)}: peak energy $E_{\rm p}$. 
		}
 		\label{grb060124:multispectra}
	\end{figure}

Our results for the best fitting model in each time interval 
are reported in Table~\ref{grb060124:multispectab}
and illustrated in Fig.~\ref{grb060124:multispectra}. 
At this time resolution, the photon index generally 
increases (softens) throughout the prompt phase (Fig.~\ref{grb060124:multispectra}c). 
The precursor is significantly softer than the main burst prompt emission. 
However, we note that in the 
X-ray range, where a finer time sampling is available for spectroscopy, the 
hardness ratio tracks the count rate on a fine scale (see Fig.~\ref{grb060124:xrtwtlc}c).
Therefore, the emission is softer when its intensity is lower 
(also, see Sect.~\ref{grb060124:xrtwtspec}).
There is marginal evidence for variations in the intrinsic column density,
with a $\sim 3$-$\sigma$  decrease between the third and fourth data point
(Fig.~\ref{grb060124:multispectra}b).
We note that the break energy lies outside the XRT band even in the 
last bin (Fig.~\ref{grb060124:multispectra}d).

For comparison, we considered the Konus spectrum around the Konus 
trigger ($t=567.3$--575.5) as representative of the most intense 
part of the burst. Fitting 
this spectrum with a {\tt GRBM} model, we obtain 
$\Gamma_1=1.13^{+0.15}_{-0.17}$, $\Gamma_2=2.12^{+0.55}_{-0.20}$
and $E_0=307_{-110}^{+190}$ ($\chi^2_{\rm red}=0.84$, 61 d.o.f.).
We derive $E_{\rm p, KW}=268_{-64}^{+99}$\,keV.

We also fitted the total spectra 
($t=301.2$--811.2\,s, last line in Table~\ref{grb060124:multispectab}). 
Based on this last fit, we derive an observed and rest-frame peak energy 
of $E_{\rm p}=193_{-39}^{+78}$\,keV and 
$E'_{\rm p}=636_{-129}^{+257}$\,keV, respectively.
We also calculated integrated quantities, 
such as the fluence $\mathcal{F}$, the peak photon and energy flux 
in several energy bands. 
These are reported in Table~\ref{grb060124:fluences}. 

The derived value of the isotropic energy, for a time integration 
of 510\,s in the source 1--10$^4$\,keV energy range, is   
$E_{\rm iso} = (4.2\pm0.5)\times 10^{53}$ erg.
With these values, this burst is consistent with the 
$E_{\rm p}$-$E_{\rm iso}$ relation  \citep{Amati2002}.  
The rest-frame isotropic luminosity in the 30--10000 keV range is 
$L_{\rm p,iso}=(1.1 \pm 0.1) \times 10^{53}$ erg s$^{-1}$.
This value is also consistent with the $E_{\rm p}$-$L_{\rm p,iso}$ 
relation \citep{Yonetoku2004}.

	\begin{figure}%%%%%%%%%%%%%%%%%%%%%%%%%%%%%%%%%%%%%%%%%%%%%%%%%%%%%%%%%%%%
		\resizebox{\hsize}{!}{\includegraphics[angle=270]{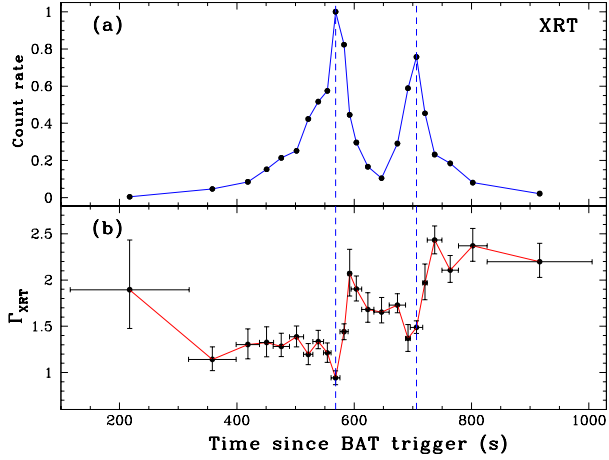}}
  		\caption{Time-resolved spectroscopy of the prompt emission of GRB~060124 with XRT/WT data. 
		{\bf (a)}: Normalized XRT count rate. 
		{\bf (b)}: Time evolution of the photon index. 
		The vertical dashed lines mark the position of the peaks.
		}
 		\label{grb060124:23bins}
	\end{figure}

	%%%%%%%%%%%%%%%%%%%%%%%%%%%%%%%%%%%%%%%%%%%%%%%%%%%%%%%%%%%%%%%%
	\subsection{Spectral evolution of the prompt phase with XRT/WT data\label{grb060124:xrtwtspec}} 
	%%%%%%%%%%%%%%%%%%%%%%%%%%%%%%%%%%%%%%%%%%%%%%%%%%%%%%%%%%%%%%%%

To study the spectral evolution during the prompt phase in more detail than possible with 
BAT or Konus data, we created 23 time intervals that covered the prompt phase, 
and extracted XRT/WT spectra in the regions defined above.
We fit them in the 0.5--10\,keV band, with an 
absorbed power law model, with the two absorption components  
described in Sect.~\ref{grb060124:multispecfits}.
The results are shown in Fig.~\ref{grb060124:23bins}. 
As Fig.~\ref{grb060124:xrtwtlc}c also shows, there is strong spectral evolution during the
prompt phase. In particular, the photon index has a minimum 
(i.e.\ the spectrum becomes harder) at the peaks of emission.
Also, we note that the
photon index achieves a plateau following each peak in the light curve,
with each plateau becoming successively softer.  During this time interval,
the photon index increases by about 1, reaching a value consistent with the
later afterglow spectrum by the end of the last peak.

	%%%%%%%%%%%%%%%%%%%%%%%%%%%%%%%%%%%%%%%%%%%%%%%%%%%%%%%%%%%%%%%%
	\subsection{XRT/PC spectrum of the afterglow\label{grb060124:xrtmeanspec}} 
	%%%%%%%%%%%%%%%%%%%%%%%%%%%%%%%%%%%%%%%%%%%%%%%%%%%%%%%%%%%%%%%%

\begin{figure}%%%%%%%%%%%%%%%%%%%%%%%%%%%%%%%%%%%%%%%%%%%%%%%%%%%%%%%%%%%%
		\resizebox{\hsize}{!}{\includegraphics[angle=270]{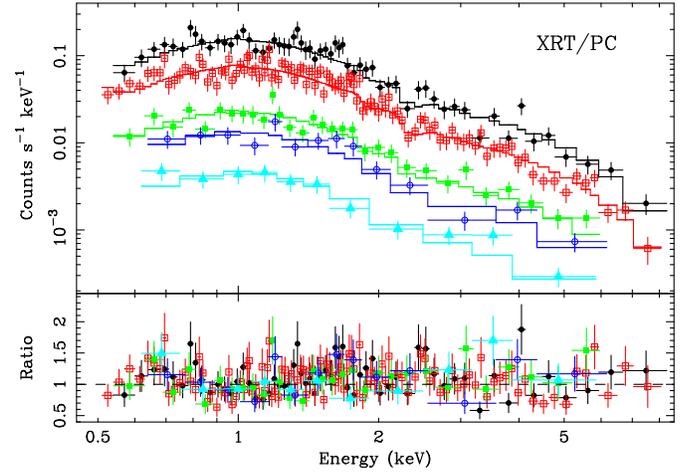}}
  		\caption{Spectroscopy of the afterglow emission of GRB~060124 with XRT/PC data. 
		{\bf Top:} PC data fit with an absorbed power law model. 
		The normalizations are relative to sequence 001, and are  
		0.29 (002), 0.09 (003), 0.05 (004), and 0.02 (005).
		{\bf Bottom:} Ratio of the best-fit models to the data.
		}
 		\label{grb060124:xrtpcspec}
	\end{figure}

The only spectroscopic information on the pure afterglow comes from the XRT/PC data.
We performed a fit of the PC data in the 0.5--10\,keV band, 
which were rebinned with a minimum of 20 counts per energy bin 
to allow $\chi^2$ fitting within XSPEC. 
We considered an absorbed power law model, with the two absorption components  
described in Sect.~\ref{grb060124:multispecfits}.
The fits for each sequence yield consistent 
photon indices and column density at the GRB redshift, as reported in 
Table~\ref{grb060124:tab_xrtpcfits}. 
We also fitted all data sets together and obtained a photon index 
$\Gamma_{\rm X, PC}=2.06\pm0.06$, $N_{\rm H}^{\rm G}=9.90\times 10^{20}$ cm$^{-2}$, 
and a column density at the GRB redshift 
$N_{\rm H}^{z} = (1.3_{-0.4}^{+0.5}) \times 10^{22}$ cm$^{-2}$
($\chi^2_{\rm red}=1.07$; 216 d.o.f., see Fig.~\ref{grb060124:xrtpcspec}).
We confirm the presence of an excess of $N_{\rm H}^{z}$ with respect to the 
Galactic Hydrogen column along the line of sight on the order of $10^{22}$ cm$^{-2}$. 

%%%%%%%%%%%%%%%%%%%%%%%%%%%%%%%%%%%%%%%%%%%%%%%%%%%%%%%%%%%%%%%%
\subsection{Summary of the observed properties of GRB~060124\label{grb060124:summaryobs}}
%%%%%%%%%%%%%%%%%%%%%%%%%%%%%%%%%%%%%%%%%%%%%%%%%%%%%%%%%%%%%%%%
\begin{enumerate}
\item The precursor/prompt phase.
	\begin{enumerate}
	\item For GRB~060124, we observed the longest interval ($\approx 500$\,s) 
		recorded between
		a precursor and the main GRB event \citep{Lazzati2005:precursors}.
		The precursor is spectrally softer than the following peaks 
		(Table~\ref{grb060124:multispectab}).
	\item The XRT, BAT and Konus light curves (0.2--2\,000\,keV) 
		show the same overall structure consisting of three peaks.
		No time-resolved information is available from UVOT,
		but we observe that the optical behaviour during the prompt 
		emission was significantly different from the X-ray one.
	\item Based on the $T_{90}$s ($\approx 300$\,s, Table~\ref{grb060124:tabt90s}), 
		although the individual peak durations are comparable to 
		those of other long GRBs, the total duration is among
		the longest recorded by either BATSE \citep{Paciesa1999:batsecatalog} 
		or Swift. 
	\item The peaks of the prompt emission shift with the energy band, with 
		the peaks observed in the harder bands preceding the ones observed 
		in the softer bands. This lag is as much as $\sim 10$\,s 
		(between 15--100\,keV and 0.2--1\,keV).
	\item Strong spectral evolution takes place during the prompt phase.
		We observe $E_{\rm p}$ moving from higher energies to lower energies
		(Sect.~\ref{grb060124:multispecfits}, Fig.~\ref{grb060124:sed}).
		The spectral evolution follows a ``tracking'' behaviour
		(Fig.~\ref{grb060124:23bins}), with the
		spectrum being softer when the flux is higher. 
		The relative importance of the three main peaks 
		varies with the energy band; the third peak in 
		the 0.1--1\,keV light curve is actually stronger than the second one, 
		as opposed to what is observed in all the other energy bands.
	\item The peaks are much broader in the softer bands.
	\item We derive a mean rest frame peak and isotropic energy 
		$E'_{\rm p}=636_{-129}^{+257}$\,keV 
		and $E_{\rm iso} = (4.2\pm0.5)\times 10^{53}$ erg.
	\end{enumerate}
\item The afterglow phase.
	\begin{enumerate}
	\item The X-ray afterglow is modelled with a 
		broken power-law decay with indices $\alpha_1\sim1.2$, $\alpha_2\sim1.6$ 
		and a break at $t_{\rm b}\sim10^5$\,s.
	\item The optical afterglow  is modelled with 
		a power-law decay with index $\alpha_{BV} \sim 0.8$
		(the optical data are only available for $t<10^5$\,s). 
	\end{enumerate}
\end{enumerate}

%%%%%%%%%%%%%%%%%%%%%%%%%%%%%%%%%%%%%%%%%%%%%%%%%%%%%%%%%%%%%%%%
\section{Discussion\label{grb060124:discussion}} %%%%%%%%%%%%%%%
%%%%%%%%%%%%%%%%%%%%%%%%%%%%%%%%%%%%%%%%%%%%%%%%%%%%%%%%%%%%%%%%

One of the main results of Swift is the identification of a steep-decay phase 
which is observed in the early X-ray light curves of many GRBs, followed by a 
shallower phase \citep{Tagliaferri2005:nature,Nousek2005:lcvs,Obrien2006:xrtbat}. 
Quite often an erratic 
flaring behaviour is observed superimposed on these phases 
\citep{Burrows2005:flarescience,Falcone2006:050502b,Romano2006:050406}.
The most accredited explanation for the steep phase is that we are seeing 
high-latitude emission due to the termination of the central engine activity
\citep{Kumar2000b,Zhang2005b}, 
while flares are probably due to the reactivation of the central engine.
Here, for the first time, we have the chance to follow in the soft X-ray 
and in the optical/UV both the prompt phase and its transition to the afterglow phase. 

	\begin{figure}%%%%%%%%%%%%%%%%%%%%%%%%%%%%%%%%%%%%%%%%%%%%%%%%%%%%%%%%%%%%
 	 	\resizebox{\hsize}{!}{\includegraphics[angle=270]{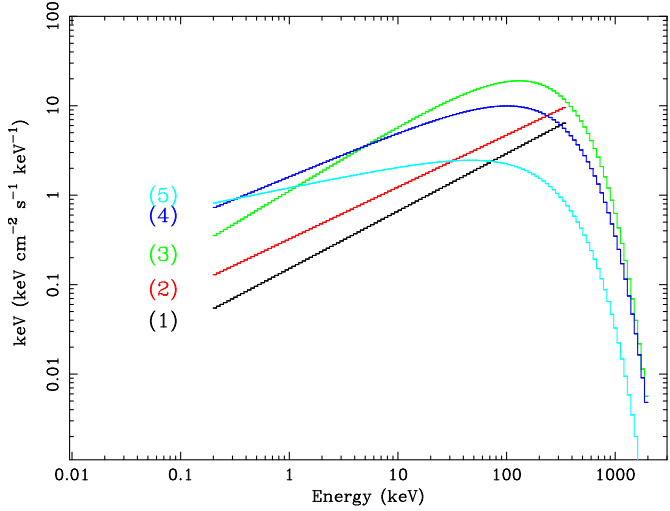}}  
		\caption{Spectral energy distribution of GRB~060124 during the prompt phase.
		SEDs (1) and (2) are derived from simultaneous fits to XRT and BAT data fits, 
		while SEDs (3)--(5) from XRT, BAT and Konus fits 
		(see Table~\ref{grb060124:multispectab}).
		}
 		\label{grb060124:sed}
	\end{figure}

The panchromatic observation of GRB~060124 during the prompt phase
provides the opportunity to study the broadband
emission of the GRB prompt emission in great detail. Time-resolved
$\nu F_\nu$ spectra were analysed for 5 temporal segments
(Fig.~\ref{grb060124:sed}), each spanning nearly 4 decades in
frequency (from the XRT to the BAT or Konus band). The
flux levels at even lower frequencies (the UVOT band) are also
measured, but at a much coarser temporal resolution.
Whenever the joint BAT-Konus data are
available (bin 3--5), a peak energy $E_{\rm p}$ can be computed. 
The time-resolved
spectral fits reveal an evolution of $E_{\rm p}$ to lower energies. 
The origin of $E_{\rm p}$
has been extensively discussed in the literature 
\citep{Meszaros1994,Tavani1996,Pilla1998,Zhang2002,Peer2004,Rees2005,Peer2005,Ryde2005}.
The suggestions range from the optically-thin characteristic synchrotron
emission frequency, to the optically-thick Comptonised spectral peak, to
the thermal peak of the fireball photosphere. In particular, there
have been questions about whether the synchrotron model can account
for the observed spectrum \citep{Lloyd2000,Ghisellini2000}, since the
standard synchrotron model requires fast cooling so that the photon
index below the $E_{\rm p}$ is expected to be $\sim 1.5$. This is
because for internal emission models (e.g.\ the internal shock model or
the internal magnetic dissipation model), the magnetic fields in the
dissipation region are so high that the electron cooling time scale is
much shorter than the dynamical time scale. The measured
photon index in the bins 2--5 are all close to this value, suggesting
that the fast-cooling synchrotron model is a plausible mechanism at
least for this burst. Lacking a temporally resolved UVOT light curve,
one cannot derive a reliable time-resolved SED extending to the UVOT
band. Nonetheless, connecting the time-averaged UVOT flux level to the
flux level in the XRT band requires a further spectral break between
the XRT and the UVOT band, which may be consistent with a cooling
break or a synchrotron self-absorption break.

Both the spectral peak energy $E_{\rm p}$ and the spectral indices
evolve with time.  This is likely 
due to the complicated interplay between particle heating and cooling
as well as the emission contributions from many emission units
(e.g. internal shocks). One interesting property is that the second
main peak (bin 5, $E_{\rm p} \sim 28$ keV) is significantly softer than the
first main peak (bin 4, $E_{\rm p} \sim 237$ keV). Within the internal
dissipation models, one has $E_{\rm p} \propto  L^{1/2} r^{-1} \propto
L^{1/2} \gamma^{-2} (\delta t)^{-1}$ \citep{Zhang2002}, 
where $L$ is the wind luminosity of the GRB
outflow, $r$ is the emission radius, $\gamma$ is the bulk Lorentz
factor, and $\delta t$ is the time variability from the central
engine. The second part of proportionality is valid specifically for
the internal shock model. 
The 15--150 keV
fluence of the second main peak $[(1.05
\pm 0.11) \times 10^{-6}~{\rm erg~cm^{-2}}]$ is about one order of
magnitude lower than that of the first main peak $[(1.25 \pm 0.03)
\times 10^{-5}~{\rm erg~cm^{-2}}]$. Given the similar durations of the
two episodes, the luminosity of the second epoch is lower by
a factor of $\sim 10$. This would account for a factor of 3 reduction
of the $E_{\rm p}$. The data then require a larger emission radius
(by a factor of $\sim$ 3). In the internal shock scenario, this
requires that the bulk Lorentz factor is somewhat higher. This may be
consistent with the idea that the environment of the burst becomes 
cleaner with time as the initial jet has loaded most of the baryons
along the path. 
This would be the first piece of evidence of an increase of the 
Lorentz factor during the prompt phase.

The fact that the pulses in the softer band tend to be broader is
related to the temporal evolution of the spectrum. If the electron
spectral index does not evolve with time (as is usually assumed
in the afterglow modeling), the rising/falling time scales of the
light curves should be similar in different energy bands. However, if
during the acceleration processes the electron spectrum hardens with
time, one can get a wider, soft wing with respect to a narrow, hard
spike. The temporal decay slopes after the peaks are shallower
than $2$. This suggests that the decay is not due to the adiabatic
cooling of the emission region. Rather, electron heating may be still
going on but with a reduced rate. If one assumes that the electron
spectral index softens with time during the decay, one can account for
a longer soft tail with respect to the narrow hard spike. 
This is supported by our finding 
that the photon indices are systematically smaller near the peaks
(Fig.~\ref{grb060124:23bins}). 

The super-long duration of the burst suggests that the GRB central
engine is active for hundreds of seconds. This has become a general
feature of both long and short GRBs in view of the commonly detected
X-ray flares hundreds of seconds after the burst trigger, which are
generally interpreted as late central engine activities 
\citep{Burrows2005:flarescience,Zhang2005b,Romano2006:050406,
Falcone2006:050502b,Barthelmy2005:nature,Campana2006:050724,
Liang2006:curvature}.   
The second main peak (excluding the precursor)
would have been categorized as an X-ray flare had it been even
softer. The ``precursor'' emission leads the main peak by about 500\,s. 
This is inconsistent with precursor models that
involve an erupting fireball from a massive stellar envelope, which
predicts a precursor leading the main episode by several seconds
\citep{Ramirez-Ruiz2002}. A straightforward possibility is that the   
precursor is also due to early central engine activity. Suggestions
to produce episodic, erratic central engine activity include
fragmentation of the collapsing star \citep{King2005} or the        
accretion disk \citep{Perna2006}, a magnetic barrier modulated      
accretion flow \citep{Proga2006}, or the magnetic activity of a     
massive neutron star central engine \citep{Dai2006}. An interesting  
question for these models would be how to generate a weak precursor
$570/(1+z) =173$\,s before the main episode.

The origin of the optical emission during the
prompt phase is uncertain from the available data. 
However, a tracking behaviour would require timescales of optical emission
in the order of $\la 10$\,s, which is short, compared with 
GRB~990123 \citep{Akerlof1999}. 
Lacking a clear
tracking behaviour with the prompt gamma-ray and X-ray light curves
suggests that it may come from a different emission site, e.g.\ the
reverse shock \citep{Meszaros1997,Meszaros1999:990123,Sari1999:990123}.
Nonetheless,
the level of the prompt optical emission clearly rules out the
presence of a prominent reverse shock component as discovered in GRB~990123 
\citep{Akerlof1999} and \object{GRB~021211} 
\citep{Fox2003,Li2003}, 
which generally requires that the magnetic fields in the reverse shock 
region are significantly stronger than that in the forward shock region
\citep{Zhang2003,Fan2002,Kumar2003}, and that
the fireball has comparable energy in baryonic materials 
and magnetic fields (i.e. $\sigma \sim 1$, \citealt{Zhang2005}).
The contribution of the reverse shock could be less significant when
$\sigma$ is low (baryonic) or very high 
\citep[Poynting-flux-dominated, ][]{Zhang2005}. In particular, various
physical processes could lead to the suppression of the reverse shock
emission even if the fireball is baryonic
\citep[e.g.][]{Kobayashi2000,Nakar2004,Kobayashi2005}. 
As a result, although the
insignificant reverse shock emission is consistent with a Poynting
flux dominated GRB model \citep[e.g.][]{Lyutikov2003,Roming2005}, 
the baryonic GRB model is not disfavoured by the data.

The early part of the afterglow ($t < t_{\rm b}$) can be explained well in
terms of the standard model \citep{SariNarayanPiran1998:ags}. In this case, the
optical and X-ray data constrain the cooling frequency $\nu_{\rm c}$ between
the optical and X-ray bands. In fact, the X-ray and optical spectral indices 
$\beta_{\rm X} = 1.1 \pm 0.10$ and $\beta_{\rm opt} = 0.5 \pm 0.3$ differ by
$\approx 0.5$, and constrain $\nu_{\rm c} \sim 2 \times 10^{16}$~Hz. The
temporal behaviour is fully consistent with this interpretation: both the
optical and X-ray decay slopes are consistent with the model prediction:
$\alpha_{\rm X} \approx 3\beta_{\rm X}/2 - 1/2 = 1.15 \pm 0.10$ (valid for $\nu
> \nu_{\rm c}$) and $\alpha_{\rm opt} \approx 3\beta_{\rm opt}/2 = 0.75 \pm
0.4$ (valid for $\nu < \nu_{\rm c}$). The latter equation is valid only for a
homogeneous external medium (the wind model would require a much steeper
$\alpha_{\rm opt} \approx 1.3$). The electron distribution index is $p = 2\beta_{\rm
X} = 2.2 \pm 0.15$, which is in agreement with the shock theory.

The X-ray decay slope after the break is too shallow for the break being due to
a standard, sideways-expanding jet, in which case $\alpha_{\rm X} = p > 2$ would
be expected \citep{Rhoads1999:beaming}. The break magnitude $\Delta\alpha_{\rm X} = 0.39
\pm 0.08$ may be consistent with a jet which propagates in a wind environment
suffering no sideways expansion (for which $\Delta\alpha = 0.5$). However, the
wind model is ruled out by the early-time optical data. A solution which
satisfies all the avaliable constraints is a structured jet having a
homogeneous core surrounded by power-law wings with energy profile ${\rm
d}E/{\rm d}\Omega \propto \vartheta^{-q}$ \citep{Panaitescu2005:jets}. The
break magnitude is dependent upon the index $q$ (and on the location of
$\nu_{\rm c}$). With the observed values, we infer $q \approx 0.85$. The break
magnitude below the cooling frequency (i.e. in the optical) is slightly
different: $\Delta\alpha_{\rm opt} = 0.47$. This difference, however, is quite small and
may be easily missed.

It is interesting to compare the expected break time according to the
$E_{\rm \gamma, iso}$-$E_{\rm p}$-$t_{\rm b}$ relation 
\citep{Ghirlanda2005b,Liang2005}, where $t_{\rm b}$
is the temporal break identified in optical data. 
According to this relation one would expect a break at 
$t_{\rm b} = (2.1\pm1.2)\times 10^5$\,s, 
for $E_{\gamma, {\rm iso}} = (4.2\pm0.5) \times 10^{53}$ 
erg and $E'_{\rm p} = 636_{-129}^{+257}$ keV,
which is broadly consistent with the observed break time $10^5$\,s. 
According to the Amati relation \citep{Amati2002,Amati2006}, the
expected intrinsic spectral break is $E'_{\rm p}({\rm Amati}) = 583$ keV. 
This is generally consistent with the measured value $E'_{\rm p} \sim
193_{-39}^{+78} (1+z) \sim 636_{-129}^{+257}$ keV.

%%%%%%%%%%%%%%%%%%%%%%%%%%%%%%%%%%%%%%%%%%%%%%%%%%%%%%%%%%%%%%%%
\section{Summary and conclusions\label{grb060124:conclusions}} %%%%%%%%%%%%%
%%%%%%%%%%%%%%%%%%%%%%%%%%%%%%%%%%%%%%%%%%%%%%%%%%%%%%%%%%%%%%%%

We have presented observations of GRB~060124, the first event for which 
both the prompt and the afterglow emission could be observed simultaneously 
and in their entirety by the three Swift instruments. 
Since BAT triggered on a precursor $\sim 570$\,s before the main burst peak,
Swift repointed the narrow field instruments to the burst position 
$\sim 350$\,s {\it before} the main burst occurred. 
GRB~060124 also triggered Konus-Wind, which observed the prompt 
emission up to $\sim $MeV energies.
Thanks to these exceptional circumstances, we could report on the temporal 
and spectral properties of the prompt emission in the optical, X-ray and 
gamma-ray ranges, as well as on the optical and X-ray properties of the 
afterglow. 

We find that while the X-ray emission (0.2--10~keV) clearly tracks the
gamma-ray burst, the optical component follows a different pattern,
likely indicating a different origin, possibly the onset of external
shocks.
The prompt GRB spectrum shows significant spectral evolution,
with both the peak energy and the spectral index varying. As observed in
several long GRBs, significant lags are measured between the hard- and
low-energy components, showing that this behaviour extends over 3
decades in energy.
The GRB peaks are shown to be much broader at soft energies. 
This is related to the temporal evolution of the spectrum, and can be
accounted for by assuming the electron spectral index softened with
time.
The burst energy ($E_{\rm iso} \sim 4.2\times 10^{53}$ erg) 
and average  rest-frame peak energy $E'_{\rm p}=636_{-129}^{+257}$\,keV) 
make GRB\,010624 consistent with the Amati relation.
The X-ray afterglow is  characterized by a 
decay presenting a break at $t_{\rm b} \sim 10^5$~s, possibly due to the 
crossing of the cooling frequency $\nu_c$.

%%%%%%%%%%%%%%%%%%%%%%%%%%%%%%%%%%%%%%%%%%%%%%%%%%%%%%%%%%%%%%%%
\begin{acknowledgements}
We thank the referee for a swift and through report 
and D.\ A. Kann for useful discussions. 
This work is supported at OAB by ASI grant I/R/039/04, 
at Penn State by NASA contract NAS5-00136 and 
at the University of Leicester by PPARC. 
We gratefully acknowledge the contributions of dozens of members 
of the XRT and UVOT team at
OAB, PSU, UL, GSFC, ASDC, and MSSL and our subcontractors, 
who helped make this instruments possible.
The Konus-Wind experiment is supported by the Russian Space Agency 
contract and RFBR grant 06-02-16070.
\end{acknowledgements}
%%%%%%%%%%%%%%%%%%%%%%%%%%%%%%%%%%%%%%%%%%%%%%%%%%%%%%%%%%%%%%%%

\setcounter{table}{1}
 \begin{table*} 	
 \begin{center} 	
 \caption{Multi-wavelength $T_{90}$ and $T_{50}$.} 	
 \label{grb060124:tabt90s} 	
 \begin{tabular}{lcrr} 
 \hline 
 \hline 
 \noalign{\smallskip} 
Obs/mode & Energy range  & $T_{90}$	 & $T_{50}$ 	    \\
	&  (keV)        & (s)	 & 	(s)    \\
 \noalign{\smallskip} 
 \hline 
 \noalign{\smallskip} 
XRT/WT & 0.2--1	&	338$\pm2$	&	160$\pm2$	\\
XRT/WT & 1--4	&	302$\pm1$	&	158$\pm1$	\\
XRT/WT & 4--10	&	291$\pm1$	&	138$\pm1$	\\
BAT    & 15--25	&	321$\pm2$	&  	 60$\pm2$	\\
BAT    & 25--50	&	324$\pm2$	&	 54$\pm2$	\\
BAT    & 50--100	&	298$\pm2$	&	 50$\pm2$	\\
Konus  & 300--1160 &	 $47\pm9$	&	 $15\pm3$	\\
  \noalign{\smallskip}
  \hline
  \end{tabular}
  \end{center}
  \begin{list}{}{} 
  \item Note: errors on $T_{90}$ and $T_{50}$ are the time-resolution of the light curve used for the calculation.
	The Konus values are probably lower limits, due to low S/N effects. 
  \end{list} 
  \end{table*}

\setcounter{table}{2}
 \begin{table*}   
  \begin{center}   
 \caption{UVOT multicolour data.}   
 \label{grb060124:uvot_mag}   
 \begin{tabular}{lrlll} 
 \hline 
 \hline 
 \noalign{\smallskip} 
  Filter&T(mid)  & T(expo)   &  Magnitude$^{\mathrm{a}}$ &   Flux density$^{\mathrm{b}}$        \\
 &   (s)   &  (s)   & (mag)    \\
 \noalign{\smallskip} 
 \hline 
 \noalign{\smallskip} 
$V$ &    183.40 &  153.31 & $16.96 \pm  0.08$ & $0.522 \pm 0.038$ \\
$V$ &    632.99 &  725.22 & $16.79 \pm  0.04$ & $0.610 \pm 0.022$ \\
$B$ &  11477.47 &  139.43 & $18.60 \pm  0.12$ & $0.220 \pm 0.024$ \\
$B$ &  17915.40 &  884.89 & $19.05 \pm  0.06$ & $0.145 \pm 0.008$ \\
$B$ &  23674.52 &  884.87 & $19.51 \pm  0.08$ & $0.095 \pm 0.007$ \\
$B$ &  29314.24 &  884.85 & $19.53 \pm  0.09$ & $0.093 \pm 0.008$ \\
$V$ &  35202.82 & 1573.78 & $19.30 \pm  0.09$ & $0.060 \pm 0.005$ \\
$V$ &  41022.66 & 1575.25 & $19.44 \pm  0.10$ & $0.053 \pm 0.005$ \\
$V$ &  46782.58 & 1575.15 & $19.52 \pm  0.10$ & $0.049 \pm 0.005$ \\
$V$ &  58332.60 & 4725.17 & $19.62 \pm  0.07$ & $0.045 \pm 0.003$ \\
$V$ &  75672.93 & 4752.29 & $19.86 \pm  0.08$ & $0.036 \pm 0.003$ \\
$V$ &  99028.58 & 7947.60 & $20.15 \pm  0.08$ & $0.028 \pm 0.002$  \\
  \noalign{\smallskip} 
 \hline
  \end{tabular}
  \end{center}
  \begin{list}{}{} 
  \item[$^{\mathrm{a}}$] Vega magnitudes.
  \item[$^{\mathrm{b}}$] In units of $10^{-16}$ erg cm$^{-2}$ s$^{-1}$ $\AA^{-1}$. 
 \item Note. Errors are 1-$\sigma$ values. %Upper limts are 3-$\sigma$ limits. 
RMS errors in the photometric zero points have not been included.
	UVOT magnitudes were not corrected for Galactic extinction. 
	Mid times are measured 	since the BAT trigger.
  \end{list} 
  \end{table*}

\setcounter{table}{3}
\begin{table*}
  \begin{center}
 \caption{Multi-wavelength spectral fits of the prompt phase of GRB~060124.}
 \label{grb060124:multispectab}
 \begin{tabular}{lllccccccc}
 \hline
 \hline
 \noalign{\smallskip}
  Time/Instrument &  Model &$N_{\rm H}^{z}$ & Photon index & $E_{\rm p}$ & Normalization & $\chi^{2}_{\rm red}$ (dof)  & Start & End\\
  Selection$^{\mathrm{a}}$  &         & (10$^{22}$ cm$^{-2}$) & & (keV) & (ph cm$^{-2}$ s$^{-1}$ keV$^{-1}$) & & (s)    & (s) \\
 \noalign{\smallskip}
 \hline
 \noalign{\smallskip}
BAT Precursor 		   & PL &          		& $1.8 \pm$ 0.2 &  & (3.4$\pm$0.4)  $\times$ 10$^{-3}$(b) & 0.88 (57)& $-1.5$ &  13.5 \\
 XRT-BAT Bin 1 		   & PL & $<4.0$ & $1.17 \pm 0.07$ &   & (7.2 $\pm$ 1.1) $\times$ 10$^{-3}$(b) & 0.96 (70) & 301.2 & 361.2 \\
 XRT-BAT Bin 2 		   & PL & 2.8$_{-0.8}^{+0.9}$ 	& $1.44  \pm 0.03$ &  & (1.26 $\pm$ 0.09) $\times$ 10$^{-3}$(b) & 0.97 (137)& 361.2 & 451.2\\
 XRT-BAT-Konus Bin 3 	   & CPL& 3.5$_{-0.4}^{+0.5}$ 	& $1.36 \pm 0.02$ & 132$_{-18}^{+29}$ & (3.0 $\pm$ 0.2) $\times$  10$^{-3}$(c) & 1.09 (404)& 451.2 & 571.2 \\
 XRT-BAT-Konus Bin 4 	   & CPL& 1.6 $\pm$ 0.3 	& $1.45 \pm 0.02$ & 237$_{-69}^{+153}$ & (1.7 $\pm 0.1$) $\times$  10$^{-3}$(c) & 0.99 (431)& 571.2 & 691.2 \\
 XRT-BAT-Konus Bin 5 	   & CPL& $1.9_{-0.4}^{+0.3}$ 	& $1.78^{+0.03}_{-0.08}$ & 28$_{-8}^{+10}$ & 4.1$_{-0.5}^{+1.4}$ $\times$ 10$^{-4}$(c) & 0.83 (393)& 691.2 & 811.2\\
XRT-BAT-Konus (1--5) 	   & CPL& 1.9 $\pm$ 0.2 	& $1.48 \pm$ 0.02 & 193$_{-39}^{+78}$ & $(1.19 \pm 0.07) \times 10^{-3}$(c) & 1.06 (618)& 301.2 & 811.2 \\
 \noalign{\smallskip}
 \hline
  \end{tabular}
  \end{center}
  \begin{list}{}{}
  \item[$^{\mathrm{a}}$] See Fig.~\ref{grb060124:multispectra}.
  \item[$^{\mathrm{b}}$] Normalization ($K$) at 50 keV.
  \item[$^{\mathrm{c}}$] Normalization ($K$) at 100 keV.
  \end{list}
  \end{table*}

\setcounter{table}{4}
\begin{table*}
  \begin{center}
 \caption{Multi-wavelength energetics of the prompt phase of GRB~060124.}
 \label{grb060124:fluences}
 \begin{tabular}{cccc}
 \hline
 \hline
 \noalign{\smallskip}
 Energy band & Energy fluence & Peak photon flux & Peak energy flux\\
(keV)       & (erg cm$^{-2}$) & (ph cm$^{-2}$ s$^{-1}$) & (erg cm$^{-2}$ s$^{-1}$)\\
 \noalign{\smallskip}
 \hline
 \noalign{\smallskip}
0.5--2  & (1.26 $\pm$ 0.03) $\times$ 10$^{-6}$ & 15.8 $\pm$ 1.2 & (2.6 $\pm$ 0.3) $\times$ 10$^{-8}$\\
2--30   & (7.3 $\pm$ 0.1) $\times$ 10$^{-6}$ & 19.1 $\pm$ 1.6 & (2.7 $\pm$ 0.3) $\times$ 10$^{-7}$\\
30--400 & (1.8 $\pm$ 0.1) $\times$ 10$^{-5}$ & 7.8 $\pm$ 0.6 & (1.4 $\pm$ 0.2) $\times$ 10$^{-6}$ \\
25--50  & (3.60 $\pm$ 0.07) $\times$ 10$^{-6}$ & 3.2 $\pm$ 0.3 & (1.8 $\pm$ 0.2) $\times$ 10$^{-6}$\\
50--100 & (4.7 $\pm$ 0.1) $\times$ 10$^{-6}$ & 2.5 $\pm$ 0.2 & (2.8 $\pm$ 0.4) $\times$ 10$^{-7}$ \\
100--300 & 8.7$_{-0.6}^{+0.9}$ $\times$ 10$^{-6}$ & 2.6 $\pm$ 0.2 & (7.1 $\pm$ 0.9) $\times$ 10$^{-7}$\\
20--2000 & 2.5$_{-0.3}^{+0.4}$ $\times$ 10$^{-5}$ & 10.7 $\pm$ 0.9 & (2.2 $\pm$ 0.3) $\times$ 10$^{-6}$\\
15--150 & (1.35 $\pm$ 0.03) $\times$ 10$^{-5}$ & 9.7 $\pm$ 0.8 & (7.8 $\pm$ 1.0) $\times$ 10$^{-7}$\\
0.5--2000 & 3.2$_{-0.3}^{+0.4}$ $\times$ 10$^{-5}$ & 43.4 $\pm$ 3.5 & (2.4 $\pm$ 0.3) $\times$ 10$^{-6}$\\ 
\noalign{\smallskip}
 \hline
  \end{tabular}
  \end{center}
  \end{table*}

\setcounter{table}{5}
 \begin{table*} 	
 \begin{center} 	
 \caption{Spectral fits of the afterglow phase of GRB~060124.} 	
 \label{grb060124:tab_xrtpcfits} 	
 \begin{tabular}{llllllll} 
 \hline 
 \hline 
 \noalign{\smallskip} 
  Spectrum & Model	& $N_{\rm H}^{z}$ & Photon index & $\chi^{2}_{\rm red}$ (d.o.f.) &  Start time  & End time  	   &  Total exposure \\
 	&	& ($10^{22}$ cm$^{-2}$)  & 	 	 &		&(s since $T_0$) 	&(s tince $T_0$) & (s)   \\
 \noalign{\smallskip} 
 \hline 
 \noalign{\smallskip} 
XRT PC (001)$^{\mathrm{a}}$ & PL &$1.6_{-0.7}^{+1.1}$ &$2.14\pm0.13$  & 1.14 (56)  & 	11408	&  30726   & 5479\\
XRT PC (002)$^{\mathrm{a}}$ & PL &$1.4_{-0.5}^{+0.7}$ &$2.08\pm0.09$  & 1.10 (104) & 	34398	&  111665  & 21979\\
XRT PC (003)$^{\mathrm{a}}$ & PL &$0.7_{-0.7}^{+1.5}$ &$1.83\pm0.19$  & 0.82 (27)  & 	115809	&  197945  & 18495\\
XRT PC (004)$^{\mathrm{a}}$ & PL &$0.8_{-0.8}^{+3.1}$ &$1.95_{-0.28}^{+0.35}$ &	 1.62 (10)  & 	202206	&  272587  & 14269\\
XRT PC (mean AG)$^{\mathrm{a}}$  & PL  	&$1.3_{-0.4}^{+0.5}$ &$2.06\pm0.06$  & 1.07 (216) & 	11408	&  475924  & 92360\\ 
  \noalign{\smallskip}
  \hline
  \end{tabular}
  \end{center}
  \begin{list}{}{}
  \item[$^{\mathrm{a}}$] Model {\tt WABS*ZWABS(POWER)} with a Galactic $N_{\rm H}^{\rm G} = 9.89\times 10^{20}$\, cm$^{-2}$ 
	and free photon index and column density at the GRB redshift.
  \end{list}
  \end{table*}

%%%%%%%%%%%%%%%%%%%%%%%%%%%%%%%%%%%%%%%%%%%%%%%%%%%%%%%%%%%%%%%%
\bibliographystyle{aa}   
\bibliography{aa5071}

\begin{thebibliography}{91}
\expandafter\ifx\csname natexlab\endcsname\relax\def\natexlab#1{#1}\fi

\bibitem[{{Akerlof} {et~al.}(1999){Akerlof}, {Balsano}, {Barthelemy}, {Bloch},
  {Butterworth}, {Casperson}, {Cline}, {Fletcher}, {Frontera}, {Gisler},
  {Heise}, {Hills}, {Kehoe}, {Lee}, {Marshall}, {McKay}, {Miller}, {Piro},
  {Priedhorsky}, {Szymanski}, \& {Wren}}]{Akerlof1999}
{Akerlof}, C., {Balsano}, R., {Barthelemy}, S., {et~al.} 1999, \nat, 398, 400

\bibitem[{{Amati}(2006)}]{Amati2006}
{Amati}, L. 2006, \mnras, submitted, arXiv:astro-ph/0601553

\bibitem[{{Amati} {et~al.}(2002){Amati}, {Frontera}, {Tavani}, {in't Zand},
  {Antonelli}, {Costa}, {Feroci}, {Guidorzi}, {Heise}, {Masetti}, {Montanari},
  {Nicastro}, {Palazzi}, {Pian}, {Piro}, \& {Soffitta}}]{Amati2002}
{Amati}, L., {Frontera}, F., {Tavani}, M., {et~al.} 2002, \aap, 390, 81

\bibitem[{{Aptekar} {et~al.}(1995){Aptekar}, {Frederiks}, {Golenetskii},
  {Ilynskii}, {Mazets}, {Panov}, {Sokolova}, {Terekhov}, {Sheshin}, {Cline}, \&
  {Stilwell}}]{KonusWind}
{Aptekar}, R.~L., {Frederiks}, D.~D., {Golenetskii}, S.~V., {et~al.} 1995,
  Space Science Reviews, 71, 265

\bibitem[{{Ballet}(1999)}]{Ballet99}
{Ballet}, J. 1999, \aaps, 135, 371

\bibitem[{{Band} {et~al.}(1993){Band}, {Matteson}, {Ford}, {Schaefer},
  {Palmer}, {Teegarden}, {Cline}, {Briggs}, {Paciesas}, {Pendleton}, {Fishman},
  {Kouveliotou}, {Meegan}, {Wilson}, \& {Lestrade}}]{Band93}
{Band}, D., {Matteson}, J., {Ford}, L., {et~al.} 1993, \apj, 413, 281

\bibitem[{{Barthelmy} {et~al.}(2005{\natexlab{a}}){Barthelmy}, {Barbier},
  {Cummings}, {Fenimore}, {Gehrels}, {Hullinger}, {Krimm}, {Markwardt},
  {Palmer}, {Parsons}, {Sato}, {Suzuki}, \& {Tueller}}]{BAT}
{Barthelmy}, S.~D., {Barbier}, L.~M., {Cummings}, J.~R., {et~al.}
  2005{\natexlab{a}}, Space Science Review, 120, 143

\bibitem[{{Barthelmy} {et~al.}(2005{\natexlab{b}}){Barthelmy}, {Chincarini},
  {Burrows}, {Gehrels}, {Covino}, {Moretti}, {Romano}, {O'Brien}, {Sarazin},
  {Kouveliotou}, {Goad}, {Vaughan}, {Tagliaferri}, {Zhang}, {Antonelli},
  {Campana}, {Cummings}, {D'Avanzo}, {Davies}, {Giommi}, {Grupe}, {Kaneko},
  {Kennea}, {King}, {Kobayashi}, {Melandri}, {M{\'e}sz{\'a}ros}, {Nousek},
  {Patel}, {Sakamoto}, \& {Wijers}}]{Barthelmy2005:nature}
{Barthelmy}, S.~D., {Chincarini}, G., {Burrows}, D.~N., {et~al.}
  2005{\natexlab{b}}, \nat, 438, 994

\bibitem[{{Bhatt} {et~al.}(2006){Bhatt}, {Sahu}, {Srividya}, \&
  {Chakradhari}}]{bhatt2006:gcn4597}
{Bhatt}, B.~C., {Sahu}, D.~K., {Srividya}, S., \& {Chakradhari}, N.~K. 2006,
  GCN Circulars, 4597, 1

\bibitem[{{Blake} {et~al.}(2005){Blake}, {Bloom}, {Starr}, {Falco},
  {Skrutskie}, {Fenimore}, {Duch{\^e}ne}, {Szentgyorgyi}, {Hornstein},
  {Prochaska}, {McCabe}, {Ghez}, {Konopacky}, {Stapelfeldt}, {Hurley},
  {Campbell}, {Kassis}, {Chaffee}, {Gehrels}, {Barthelmy}, {Cummings},
  {Hullinger}, {Krimm}, {Markwardt}, {Palmer}, {Parsons}, {McLean}, \&
  {Tueller}}]{Blake:041219a}
{Blake}, C.~H., {Bloom}, J.~S., {Starr}, D.~L., {et~al.} 2005, \nat, 435, 181

\bibitem[{{Bo{\"e}r} {et~al.}(2005){Bo{\"e}r}, {Atteia}, {Damerdji}, {Gendre},
  {Klotz}, \& {Stratta}}]{grb050904opt}
{Bo{\"e}r}, M., {Atteia}, J.~L., {Damerdji}, Y., {et~al.} 2005,
  {arXiv:astro-ph/051038}

\bibitem[{{Burrows} {et~al.}(2005{\natexlab{a}}){Burrows}, {Hill}, {Nousek},
  {Kennea}, {Well}, {Osborne}, {Abbey}, {Beardmore}, {Mukerjee}, \&
  {Short}}]{XRT}
{Burrows}, D.~N., {Hill}, J.~E., {Nousek}, J.~A., {et~al.} 2005{\natexlab{a}},
  Space Science Review, 120, 165

\bibitem[{{Burrows} {et~al.}(2005{\natexlab{b}}){Burrows}, {Romano}, {Falcone},
  {Kobayashi}, {Zhang}, {Moretti}, {O'Brien}, {Goad}, {Campana}, {Page},
  {Angelini}, {Barthelmy}, {Beardmore}, {Capalbi}, {Chincarini}, {Cummings},
  {Cusumano}, {Fox}, {Giommi}, {Hill}, {Kennea}, {Krimm}, {Mangano},
  {Marshall}, {M{\'e}sz{\'a}ros}, {Morris}, {Nousek}, {Osborne}, {Pagani},
  {Perri}, {Tagliaferri}, {Wells}, {Woosley}, \&
  {Gehrels}}]{Burrows2005:flarescience}
{Burrows}, D.~N., {Romano}, P., {Falcone}, A., {et~al.} 2005{\natexlab{b}},
  Science, 309, 1833

\bibitem[{{Campana} {et~al.}(2006){Campana}, {Tagliaferri}, {Lazzati},
  {Chincarini}, {Covino}, {Page}, \& {Romano}}]{Campana2006:050724}
{Campana}, S., {Tagliaferri}, G., {Lazzati}, D., {et~al.} 2006, \aap, in press

\bibitem[{{Cenko} {et~al.}(2006){Cenko}, {Berger}, \&
  {Cohen}}]{cenko2006:gcn4592}
{Cenko}, S.~B., {Berger}, E., \& {Cohen}, J. 2006, GCN Circulars, 4592, 1

\bibitem[{{Dai} {et~al.}(2006){Dai}, {Wang}, {Wu}, \& {Zhang}}]{Dai2006}
{Dai}, Z.~G., {Wang}, X.~Y., {Wu}, X.~F., \& {Zhang}, B. 2006, Science, 311,
  1127

\bibitem[{{Dickey} \& {Lockman}(1990)}]{DL90}
{Dickey}, J.~M. \& {Lockman}, F.~J. 1990, \araa, 28, 215

\bibitem[{{Falcone} {et~al.}(2006){Falcone}, {Burrows}, {Lazzati}, {Campana},
  {Kobayashi}, {Zhang}, {M{\'e}sz{\'a}ros}, {Page}, {Kennea}, {Romano},
  {Pagani}, {Angelini}, {Beardmore}, {Capalbi}, {Chincarini}, {Cusumano},
  {Giommi}, {Goad}, {Godet}, {Grupe}, {Hill}, {La Parola}, {Mangano},
  {Moretti}, {Nousek}, {O'Brien}, {Osborne}, {Perri}, {Tagliaferri}, {Wells},
  \& {Gehrels}}]{Falcone2006:050502b}
{Falcone}, A.~D., {Burrows}, D.~N., {Lazzati}, D., {et~al.} 2006, \apj, 641,
  1010

\bibitem[{{Fan} {et~al.}(2002){Fan}, {Dai}, {Huang}, \& {Lu}}]{Fan2002}
{Fan}, Y.-Z., {Dai}, Z.-G., {Huang}, Y.-F., \& {Lu}, T. 2002, Chinese Journal
  of Astronony and Astrophysics, 2, 449

\bibitem[{{Fenimore} {et~al.}(2006){Fenimore}, {Barbier}, {Barthelmy},
  {Cummings}, {Gehrels}, {Hullinger}, {Krimm}, {Markwardt}, {McMahon},
  {Palmer}, {Parsons}, {Sakamoto}, {Sato}, {Tueller}, \&
  {White}}]{fenimore2006:gcn4586}
{Fenimore}, E., {Barbier}, L., {Barthelmy}, S., {et~al.} 2006, GCN Circulars,
  4586, 1

\bibitem[{{Ford} {et~al.}(1995){Ford}, {Band}, {Matteson}, {Briggs},
  {Pendleton}, {Preece}, {Paciesas}, {Teegarden}, {Palmer}, {Schaefer},
  {Cline}, {Fishman}, {Kouveliotou}, {Meegan}, {Wilson}, \&
  {Lestrade}}]{Ford1995:softhard}
{Ford}, L.~A., {Band}, D.~L., {Matteson}, J.~L., {et~al.} 1995, \apj, 439, 307

\bibitem[{{Fox} {et~al.}(2003){Fox}, {Price}, {Soderberg}, {Berger},
  {Kulkarni}, {Sari}, {Frail}, {Harrison}, {Yost}, {Matthews}, {Peterson},
  {Tanaka}, {Christiansen}, \& {Moriarty-Schieven}}]{Fox2003}
{Fox}, D.~W., {Price}, P.~A., {Soderberg}, A.~M., {et~al.} 2003, \apjl, 586, L5

\bibitem[{{Gehrels} {et~al.}(2004){Gehrels}, {Chincarini}, {Giommi}, {Mason},
  {Nousek}, {Wells}, {White}, {Barthelmy}, {Burrows}, {Cominsky}, {Hurley},
  {Marshall}, {M{\' e}sz{\' a}ros}, {Roming}, {Angelini}, {Barbier}, {Belloni},
  {Campana}, {Caraveo}, {Chester}, {Citterio}, {Cline}, {Cropper}, {Cummings},
  {Dean}, {Feigelson}, {Fenimore}, {Frail}, {Fruchter}, {Garmire}, {Gendreau},
  {Ghisellini}, {Greiner}, {Hill}, {Hunsberger}, {Krimm}, {Kulkarni}, {Kumar},
  {Lebrun}, {Lloyd-Ronning}, {Markwardt}, {Mattson}, {Mushotzky}, {Norris},
  {Osborne}, {Paczynski}, {Palmer}, {Park}, {Parsons}, {Paul}, {Rees},
  {Reynolds}, {Rhoads}, {Sasseen}, {Schaefer}, {Short}, {Smale}, {Smith},
  {Stella}, {Tagliaferri}, {Takahashi}, {Tashiro}, {Townsley}, {Tueller},
  {Turner}, {Vietri}, {Voges}, {Ward}, {Willingale}, {Zerbi}, \&
  {Zhang}}]{Gehrels04}
{Gehrels}, N., {Chincarini}, G., {Giommi}, P., {et~al.} 2004, \apj, 611, 1005

\bibitem[{{Ghirlanda} {et~al.}(2005){Ghirlanda}, {Ghisellini}, \&
  {Firmani}}]{Ghirlanda2005b}
{Ghirlanda}, G., {Ghisellini}, G., \& {Firmani}, C. 2005, \mnras, 361, L10

\bibitem[{{Ghisellini} {et~al.}(2000){Ghisellini}, {Celotti}, \&
  {Lazzati}}]{Ghisellini2000}
{Ghisellini}, G., {Celotti}, A., \& {Lazzati}, D. 2000, \mnras, 313, L1

\bibitem[{{Golenetskii} {et~al.}(2006){Golenetskii}, {Aptekar}, {Mazets},
  {Pal'shin}, {Frederiks}, \& {Cline}}]{golenetskii2006:gcn4599}
{Golenetskii}, S., {Aptekar}, R., {Mazets}, E., {et~al.} 2006, GCN Circulars,
  4599, 1

\bibitem[{{Greco} {et~al.}(2006){Greco}, {Bartolini}, {Guarnieri}, {Piccioni},
  {Pizzichini}, {Silvotti}, {Nanni}, {Terra}, \& {Bruni}}]{greco2006:gcn4605}
{Greco}, G., {Bartolini}, C., {Guarnieri}, A., {et~al.} 2006, GCN Circulars,
  4605, 1

\bibitem[{{Hill} {et~al.}(2004){Hill}, {Burrows}, {Nousek}, {Abbey}, {Ambrosi},
  {Br{\"a}uninger}, {Burkert}, {Campana}, {Cheruvu}, {Cusumano}, {Freyberg},
  {Hartner}, {Klar}, {Mangels}, {Moretti}, {Mori}, {Morris}, {Short},
  {Tagliaferri}, {Watson}, {Wood}, \& {Wells}}]{Hill04}
{Hill}, J.~E., {Burrows}, D.~N., {Nousek}, J.~A., {et~al.} 2004, in X-Ray and
  Gamma-Ray Instrumentation for Astronomy XIII. Edited by Flanagan, Kathryn A.;
  Siegmund, Oswald H. W. Proceedings of the SPIE, Volume 5165, pp. 217-231
  (2004)., ed. K.~A. {Flanagan} \& O.~H.~W. {Siegmund}, 217--231

\bibitem[{{Holland} {et~al.}(2006){Holland}, {Barthelmy}, {Burrows}, {Gehrels},
  {Hunsberger}, {Kennea}, {Parola}, {Markwardt}, {Page}, {Palmer}, \&
  {Sakamoto}}]{holland2006:gcn4570}
{Holland}, S.~T., {Barthelmy}, S., {Burrows}, D.~N., {et~al.} 2006, GCN
  Circulars, 4570, 1

\bibitem[{{Kalberla} {et~al.}(2005){Kalberla}, {Burton}, {Hartmann}, {Arnal},
  {Bajaja}, {Morras}, \& {P{\"o}ppel}}]{LABS}
{Kalberla}, P.~M.~W., {Burton}, W.~B., {Hartmann}, D., {et~al.} 2005, \aap,
  440, 775

\bibitem[{{Kann}(2006)}]{kann2006:gcn4574}
{Kann}, D.~A. 2006, GCN Circulars, 4574, 1

\bibitem[{{King} {et~al.}(2005){King}, {O'Brien}, {Goad}, {Osborne}, {Olsson},
  \& {Page}}]{King2005}
{King}, A., {O'Brien}, P.~T., {Goad}, M.~R., {et~al.} 2005, \apjl, 630, L113

\bibitem[{{Klotz} {et~al.}(2006){Klotz}, {Boer}, \&
  {Atteia}}]{klotz2006:gcn4581}
{Klotz}, A., {Boer}, M., \& {Atteia}, J.~L. 2006, GCN Circulars, 4581, 1

\bibitem[{{Kobayashi}(2000)}]{Kobayashi2000}
{Kobayashi}, S. 2000, \apj, 545, 807

\bibitem[{{Kobayashi} {et~al.}(2005){Kobayashi}, {Zhang}, {M{\'e}sz{\'a}ros},
  \& {Burrows}}]{Kobayashi2005}
{Kobayashi}, S., {Zhang}, B., {M{\'e}sz{\'a}ros}, P., \& {Burrows}, D.~N. 2005,
  {\apjl, submitted, arXiv:astro-ph/0506157}

\bibitem[{{Kumar} \& {Panaitescu}(2000)}]{Kumar2000b}
{Kumar}, P. \& {Panaitescu}, A. 2000, \apjl, 541, L51

\bibitem[{{Kumar} \& {Panaitescu}(2003)}]{Kumar2003}
{Kumar}, P. \& {Panaitescu}, A. 2003, \mnras, 346, 905

\bibitem[{{Lamb} {et~al.}(2006){Lamb}, {Ricker}, {Atteia}, {Kawai}, \&
  {Woosley}}]{lamb2006:gcn4601}
{Lamb}, D., {Ricker}, G., {Atteia}, J., {Kawai}, N., \& {Woosley}, S. 2006, GCN
  Circulars, 4601, 1

\bibitem[{{Lazzati}(2005)}]{Lazzati2005:precursors}
{Lazzati}, D. 2005, \mnras, 357, 722

\bibitem[{{Li} {et~al.}(2003){Li}, {Filippenko}, {Chornock}, \& {Jha}}]{Li2003}
{Li}, W., {Filippenko}, A.~V., {Chornock}, R., \& {Jha}, S. 2003, \apjl, 586,
  L9

\bibitem[{{Liang} \& {Zhang}(2005)}]{Liang2005}
{Liang}, E. \& {Zhang}, B. 2005, \apj, 633, 611

\bibitem[{{Liang} {et~al.}(2006){Liang}, {Zhang}, {O'Brien}, {Willingale},
  {Angelini}, {Burrows}, {Campana}, {Chincarini}, {Falcone}, {Gehrels}, {Goad},
  {Grupe}, {Kobayashi}, {M{\'e}sz{\'a}ros}, {Nousek}, {Osborne}, {Page}, \&
  {Tagliaferri}}]{Liang2006:curvature}
{Liang}, E., {Zhang}, B., {O'Brien}, P.~T., {et~al.} 2006, \apj, submitted,
  astro-ph/0602142

\bibitem[{{Lipunov} {et~al.}(2006){Lipunov}, {Kornilov}, {Kuvshinov},
  {Tyurina}, {Belinski}, {Gorbovskoy}, {Krylov}, {Borisov}, {Sankovich},
  {Antipov}, \& {Vladimirov}}]{lipunov2006:gcn4572}
{Lipunov}, V., {Kornilov}, V., {Kuvshinov}, D., {et~al.} 2006, GCN Circulars,
  4572, 1

\bibitem[{{Lloyd} \& {Petrosian}(2000)}]{Lloyd2000}
{Lloyd}, N.~M. \& {Petrosian}, V. 2000, \apj, 543, 722

\bibitem[{{Lyutikov} \& {Blandford}(2003)}]{Lyutikov2003}
{Lyutikov}, M. \& {Blandford}, R. 2003, {arXiv:astro-ph/0312347}

\bibitem[{{Mangano} {et~al.}(2006){Mangano}, {Cusumano}, {Parola}, {Mineo}, \&
  {Burrows}}]{mangano2006:gcn4578}
{Mangano}, V., {Cusumano}, G., {Parola}, V.~L., {Mineo}, T., \& {Burrows},
  D.~N. 2006, GCN Circulars, 4578, 1

\bibitem[{{Masetti} {et~al.}(2006){Masetti}, {Palazzi}, {Maiorano}, {Pian},
  {Giro}, {Bonoli}, \& {Malesani}}]{masetti2006:gcn4587}
{Masetti}, N., {Palazzi}, E., {Maiorano}, E., {et~al.} 2006, GCN Circulars,
  4587, 1

\bibitem[{{M{\'e}sz{\'a}ros} \& {Rees}(1997)}]{Meszaros1997}
{M{\'e}sz{\'a}ros}, P. \& {Rees}, M.~J. 1997, \apj, 476, 232

\bibitem[{{M{\'e}sz{\'a}ros} \& {Rees}(1999)}]{Meszaros1999:990123}
{M{\'e}sz{\'a}ros}, P. \& {Rees}, M.~J. 1999, \mnras, 306, L39

\bibitem[{{M{\'e}sz{\'a}ros} {et~al.}(1994){M{\'e}sz{\'a}ros}, {Rees}, \&
  {Papathanassiou}}]{Meszaros1994}
{M{\'e}sz{\'a}ros}, P., {Rees}, M.~J., \& {Papathanassiou}, H. 1994, \apj, 432,
  181

\bibitem[{{Mirabal} \& {Halpern}(2006)}]{mirabal2006:gcn4591}
{Mirabal}, N. \& {Halpern}, J.~P. 2006, GCN Circulars, 4591, 1

\bibitem[{{Misra}(2006)}]{misra2006:gcn4589}
{Misra}, K. 2006, GCN Circulars, 4589, 1

\bibitem[{{Nakar} \& {Piran}(2004)}]{Nakar2004}
{Nakar}, E. \& {Piran}, T. 2004, \mnras, 353, 647

\bibitem[{{Nousek} {et~al.}(2006){Nousek}, {Kouveliotou}, {Grupe}, {Page},
  {Granot}, {Ramirez-Ruiz}, {Patel}, {Burrows}, {Mangano}, {Barthelmy},
  {Beardmore}, {Campana}, {Capalbi}, {Chincarini}, {Cusumano}, {Falcone},
  {Gehrels}, {Giommi}, {Goad}, {Godet}, {Hurkett}, {Kennea}, {Moretti},
  {O'Brien}, {Osborne}, {Romano}, {Tagliaferri}, \& {Wells}}]{Nousek2005:lcvs}
{Nousek}, J.~A., {Kouveliotou}, C., {Grupe}, D., {et~al.} 2006, \apj, in press,
  arXiv:astro-ph/0508332

\bibitem[{{O'Brien} {et~al.}(2006){O'Brien}, {Willingale}, {Osborne}, {Goad},
  {Page}, {Vaughan}, {Rol}, {Beardmore}, {Godet}, {Hurkett}, {Wells}, {Zhang},
  {Kobayash}, {Burrows}, {Nousek}, {Kennea}, {Falcone}, {Grupe}, {Gehrels},
  {Barthelmy}, {Cannizzo}, {Cummings}, {Hill}, {Krimm}, {Chincarini},
  {Tagliaferri}, {Campana}, {Moretti}, {Giommi}, {Perri}, {Mangano}, \& {La
  Parola}}]{Obrien2006:xrtbat}
{O'Brien}, P.~T., {Willingale}, R., {Osborne}, J., {et~al.} 2006, \apj,
  sumbitted, astro-ph/0601125

\bibitem[{{Paciesas} {et~al.}(1999){Paciesas}, {Meegan}, {Pendleton}, {Briggs},
  {Kouveliotou}, {Koshut}, {Lestrade}, {McCollough}, {Brainerd}, {Hakkila},
  {Henze}, {Preece}, {Connaughton}, {Kippen}, {Mallozzi}, {Fishman},
  {Richardson}, \& {Sahi}}]{Paciesa1999:batsecatalog}
{Paciesas}, W.~S., {Meegan}, C.~A., {Pendleton}, G.~N., {et~al.} 1999, \apjs,
  122, 465

\bibitem[{{Panaitescu}(2005)}]{Panaitescu2005:jets}
{Panaitescu}, A. 2005, \mnras, 362, 921

\bibitem[{{Pe'er} {et~al.}(2005){Pe'er}, {M{\'e}sz{\'a}ros}, \&
  {Rees}}]{Peer2005}
{Pe'er}, A., {M{\'e}sz{\'a}ros}, P., \& {Rees}, M.~J. 2005, \apj, 635, 476

\bibitem[{{Pe'er} \& {Waxman}(2004)}]{Peer2004}
{Pe'er}, A. \& {Waxman}, E. 2004, \apj, 613, 448

\bibitem[{{Perna} {et~al.}(2006){Perna}, {Armitage}, \& {Zhang}}]{Perna2006}
{Perna}, R., {Armitage}, P.~J., \& {Zhang}, B. 2006, \apjl, 636, L29

\bibitem[{{Pilla} \& {Loeb}(1998)}]{Pilla1998}
{Pilla}, R.~P. \& {Loeb}, A. 1998, \apjl, 494, L167

\bibitem[{{Predehl} \& {Schmitt}(1995)}]{avnh}
{Predehl}, P. \& {Schmitt}, J.~H.~M.~M. 1995, \aap, 293, 889

\bibitem[{{Prochaska} {et~al.}(2006){Prochaska}, {Foley}, {Tran}, {Bloom}, \&
  {Chen}}]{prochaska2006:gcn4593}
{Prochaska}, J.~X., {Foley}, R., {Tran}, H., {Bloom}, J.~S., \& {Chen}, H.-W.
  2006, GCN Circulars, 4593, 1

\bibitem[{{Proga} \& {Zhang}(2006)}]{Proga2006}
{Proga}, D. \& {Zhang}, B. 2006, \apj, submitted, arXiv:astro-ph/0601272

\bibitem[{{Quimby} {et~al.}(2006){Quimby}, {Rykoff}, {Yost}, {Aharonian},
  {Akerlof}, {Alatalo}, {Ashley}, {Goegues}, {Guever}, {Horns}, {Kehoe},
  {Kiziloglu}, {McKay}, {Oezel}, {Phillips}, {Schaefer}, {Smith}, {Swan},
  {Vestrand}, {Wheeler}, \& {Wren}}]{Quimby2006:050319}
{Quimby}, R.~M., {Rykoff}, E.~S., {Yost}, S.~A., {et~al.} 2006, \apj, in press,
  arXiv:astro-ph/0511421

\bibitem[{{Ramirez-Ruiz} {et~al.}(2002){Ramirez-Ruiz}, {MacFadyen}, \&
  {Lazzati}}]{Ramirez-Ruiz2002}
{Ramirez-Ruiz}, E., {MacFadyen}, A.~I., \& {Lazzati}, D. 2002, \mnras, 331, 197

\bibitem[{{Rees} \& {M{\'e}sz{\'a}ros}(2005)}]{Rees2005}
{Rees}, M.~J. \& {M{\'e}sz{\'a}ros}, P. 2005, \apj, 628, 847

\bibitem[{{Rhoads}(1999)}]{Rhoads1999:beaming}
{Rhoads}, J.~E. 1999, \apj, 525, 737

\bibitem[{{Romano} {et~al.}(2006){Romano}, {Moretti}, {Banat}, {Burrows},
  {Campana}, {Chincarini}, {Covino}, {Malesani}, {Tagliaferri}, {Kobayashi},
  {Zhang}, {Falcone}, {Angelini}, {Barthelmy}, {Beardmore}, {Capalbi},
  {Cusumano}, {Giommi}, {Goad}, {Godet}, {Grupe}, {Hill}, {Kennea}, {La
  Parola}, {Mangano}, {M{\'e}sz{\'a}ros}, {Morris}, {Nousek}, {O'Brien},
  {Osborne}, {Parsons}, {Perri}, {Pagani}, {Page}, {Wells}, \&
  {Gehrels}}]{Romano2006:050406}
{Romano}, P., {Moretti}, A., {Banat}, P.~L., {et~al.} 2006, \aap, 450, 59

\bibitem[{{Roming} {et~al.}(2005{\natexlab{a}}){Roming}, {Kennedy}, {Mason},
  {Nousek}, {Ahr}, {Bingham}, {Broos}, {Carter}, {Hancock}, {Huckle},
  {Hunsberger}, {Kawakami}, {Killough}, {Koch}, {Mclelland}, {Smith}, {Smith},
  {Soto}, {Boyd}, {Breeveld}, {Holland}, {Ivanushkina}, {Pryzby}, {Still}, \&
  {Stock}}]{UVOT}
{Roming}, P.~W.~A., {Kennedy}, T.~E., {Mason}, K.~O., {et~al.}
  2005{\natexlab{a}}, Space Science Review, 120, 95

\bibitem[{{Roming} {et~al.}(2005{\natexlab{b}}){Roming}, {Schady}, {Fox},
  {Zhang}, {Liang}, {Mason}, {Rol}, {Burrows}, {Blustin}, {Boyd}, {Brown},
  {Holland}, {McGowan}, {Landsman}, {Page}, {Rhoads}, {Rosen}, {Barthelmy},
  {Breeveld}, {Cucchiara}, {De Pasquale}, {Fenimore}, {Gehrels}, {Gronwall},
  {Grupe}, {Goad}, {Ivanushkina}, {James}, {Kennea}, {Kobayashi}, {Mangano},
  {M{\'e}sz{\'a}ros}, {Morgan}, {Nousek}, {Osborne}, {Palmer}, {Poole},
  {Still}, {Tagliaferri}, \& {Zane}}]{Roming2005}
{Roming}, P.~W.~A., {Schady}, P., {Fox}, D.~B., {et~al.} 2005{\natexlab{b}},
  {arXiv:astro-ph/0509273}

\bibitem[{{Rumyantsev} {et~al.}(2006){Rumyantsev}, {Biryukov}, \&
  {Pozanenko}}]{rumyantsev2006:gcn4609}
{Rumyantsev}, V., {Biryukov}, V., \& {Pozanenko}, A. 2006, GCN Circulars, 4609,
  1

\bibitem[{{Ryde}(2005)}]{Ryde2005}
{Ryde}, F. 2005, \apjl, 625, L95

\bibitem[{{Rykoff} {et~al.}(2006){Rykoff}, {Mangano}, {Yost}, {Sari},
  {Aharonian}, {Akerlof}, {Ashley}, {Barthelmy}, {Burrows}, {Gehrels},
  {G{\"o}{\v g}{\"u}{\c s}}, {G{\"u}ver}, {Horns}, {K{\i}z{\i}lo{\v g}lu},
  {Krimm}, {McKay}, {{\"O}zel}, {Phillips}, {Quimby}, {Rowell}, {Rujopakarn},
  {Schaefer}, {Smith}, {Swan}, {Vestrand}, {Wheeler}, {Wren}, \&
  {Yuan}}]{Rykoff2006:050801}
{Rykoff}, E.~S., {Mangano}, V., {Yost}, S.~A., {et~al.} 2006, \apjl, 638, L5

\bibitem[{{Rykoff} {et~al.}(2005){Rykoff}, {Yost}, {Krimm}, {Aharonian},
  {Akerlof}, {Alatalo}, {Ashley}, {Barthelmy}, {Gehrels}, {G{\"o}{\v g}{\"u}{\c
  s}}, {G{\"u}ver}, {Horns}, {K{\i}z{\i}lo{\v g}lu}, {McKay}, {{\"O}zel},
  {Phillips}, {Quimby}, {Rujopakarn}, {Schaefer}, {Smith}, {Swan}, {Vestrand},
  {Wheeler}, \& {Wren}}]{Rykoff2005:050401}
{Rykoff}, E.~S., {Yost}, S.~A., {Krimm}, H.~A., {et~al.} 2005, \apjl, 631, L121

\bibitem[{{Sari} \& {Piran}(1999{\natexlab{a}})}]{Sari1999:990123}
{Sari}, R. \& {Piran}, T. 1999{\natexlab{a}}, \apjl, 517, L109

\bibitem[{{Sari} \& {Piran}(1999{\natexlab{b}})}]{Sari1999:opticalflash}
{Sari}, R. \& {Piran}, T. 1999{\natexlab{b}}, \apj, 520, 641

\bibitem[{{Sari} {et~al.}(1998){Sari}, {Piran}, \&
  {Narayan}}]{SariNarayanPiran1998:ags}
{Sari}, R., {Piran}, T., \& {Narayan}, R. 1998, \apjl, 497, L17+

\bibitem[{{Schlegel} {et~al.}(1998){Schlegel}, {Finkbeiner}, \&
  {Davis}}]{Schlegelea98}
{Schlegel}, D.~J., {Finkbeiner}, D.~P., \& {Davis}, M. 1998, \apj, 500, 525

\bibitem[{{Sonoda} {et~al.}(2006){Sonoda}, {Maeno}, {Nakamura}, {Masuda}, \&
  {Yamauchi}}]{sonoda2006:gcn4576}
{Sonoda}, E., {Maeno}, S., {Nakamura}, Y., {Masuda}, S., \& {Yamauchi}, M.
  2006, GCN Circulars, 4576, 1

\bibitem[{{Tagliaferri} {et~al.}(2005){Tagliaferri}, {Goad}, {Chincarini},
  {Moretti}, {Campana}, {Burrows}, {Perri}, {Barthelmy}, {Gehrels}, {Krimm},
  {Sakamoto}, {Kumar}, {M{\'e}sz{\'a}ros}, {Kobayashi}, {Zhang}, {Angelini},
  {Banat}, {Beardmore}, {Capalbi}, {Covino}, {Cusumano}, {Giommi}, {Godet},
  {Hill}, {Kennea}, {Mangano}, {Morris}, {Nousek}, {O'Brien}, {Osborne},
  {Pagani}, {Page}, {Romano}, {Stella}, \& {Wells}}]{Tagliaferri2005:nature}
{Tagliaferri}, G., {Goad}, M., {Chincarini}, G., {et~al.} 2005, \nat, 436, 985

\bibitem[{{Tavani}(1996)}]{Tavani1996}
{Tavani}, M. 1996, Physical Review Letters, 76, 3478

\bibitem[{{Torii}(2006)}]{torii2006:gcn4596}
{Torii}, K. 2006, GCN Circulars, 4596, 1

\bibitem[{{Vaughan} {et~al.}(2006){Vaughan}, {Goad}, {Beardmore}, {O'Brien},
  {Osborne}, {Page}, {Barthelmy}, {Burrows}, {Campana}, {Cannizzo}, {Capalbi},
  {Chincarini}, {Cummings}, {Cusumano}, {Giommi}, {Godet}, {Hill}, {Kobayashi},
  {Kumar}, {La Parola}, {Levan}, {Mangano}, {M{\'e}sz{\'a}ros}, {Moretti},
  {Morris}, {Nousek}, {Pagani}, {Palmer}, {Racusin}, {Romano}, {Tagliaferri},
  {Zhang}, \& {Gehrels}}]{vaughan2006:050315}
{Vaughan}, S., {Goad}, M.~R., {Beardmore}, A.~P., {et~al.} 2006, \apj, 638, 920

\bibitem[{{Vestrand} {et~al.}(2005){Vestrand}, {Wozniak}, {Wren}, {Fenimore},
  {Sakamoto}, {White}, {Casperson}, {Davis}, {Evans}, {Galassi}, {McGowan},
  {Schier}, {Asa}, {Barthelmy}, {Cummings}, {Gehrels}, {Hullinger}, {Krimm},
  {Markwardt}, {McLean}, {Palmer}, {Parsons}, \& {Tueller}}]{Vestrand:041219a}
{Vestrand}, W.~T., {Wozniak}, P.~R., {Wren}, J.~A., {et~al.} 2005, \nat, 435,
  178

\bibitem[{{Wo{\'z}niak} {et~al.}(2005){Wo{\'z}niak}, {Vestrand}, {Wren},
  {White}, {Evans}, \& {Casperson}}]{Wozniak2005:050319}
{Wo{\'z}niak}, P.~R., {Vestrand}, W.~T., {Wren}, J.~A., {et~al.} 2005, \apjl,
  627, L13

\bibitem[{{Yonetoku} {et~al.}(2004){Yonetoku}, {Murakami}, {Nakamura},
  {Yamazaki}, {Inoue}, \& {Ioka}}]{Yonetoku2004}
{Yonetoku}, D., {Murakami}, T., {Nakamura}, T., {et~al.} 2004, \apj, 609, 935

\bibitem[{{Zhang} {et~al.}(2006){Zhang}, {Fan}, {Dyks}, {Kobayashi},
  {M{\'e}sz{\'a}ros}, {Burrows}, {Nousek}, \& {Gehrels}}]{Zhang2005b}
{Zhang}, B., {Fan}, Y.~Z., {Dyks}, J., {et~al.} 2006, {\apj, in press,
  arXiv:astro-ph/0508321}

\bibitem[{{Zhang} \& {Kobayashi}(2005)}]{Zhang2005}
{Zhang}, B. \& {Kobayashi}, S. 2005, \apj, 628, 315

\bibitem[{{Zhang} {et~al.}(2003){Zhang}, {Kobayashi}, \&
  {M{\'e}sz{\'a}ros}}]{Zhang2003}
{Zhang}, B., {Kobayashi}, S., \& {M{\'e}sz{\'a}ros}, P. 2003, \apj, 595, 950

\bibitem[{{Zhang} \& {M{\'e}sz{\'a}ros}(2002)}]{Zhang2002}
{Zhang}, B. \& {M{\'e}sz{\'a}ros}, P. 2002, \apj, 581, 1236

\end{thebibliography}

%%%%%%%%%%%%%%%%%%%%%%%%
\Online

\setcounter{table}{0}
\begin{table*} 	
 \begin{center} 	
 \caption{Observation log of GRB 060124.} 	
 \label{grb060124:tab_obs} 	
 \begin{tabular}{llllll} 
 \hline
 \hline
 \noalign{\smallskip}
Sequence   & Obs/mode  & Start time  (UT)  & End time   (UT) & Net exposure & Time since trigger   \\
           &           & (yyyy-mm-dd hh:mm:ss)  & (yyyy-mm-dd hh:mm:ss)     &(s)  & (s)       \\
 \noalign{\smallskip}
 \hline
 \noalign{\smallskip}
00178750000	&  BAT/event-by-event   & 2006-01-24 15:54:13 &	2006-01-24 15:59:55	& 342.3		& $-40$ 	\\
00178750000	&  BAT/mask-weighted LC & 2006-01-24 15:55:12 &	2006-01-24 16:14:51	& 1179.2	& 19.2	\\
00178750000	&  BAT/DPH              & 2006-01-24 15:59:54 &	2006-01-24 16:08:24	& 510		&301.2	\\
 \hline
00178750000	&	XRT/IM	&	2006-01-24 15:56:36	&	2006-01-24 15:56:38	&	2.5	&	104	\\
00178750000	&	XRT/WT	&	2006-01-24 15:56:44	&	2006-01-24 16:11:34	&	867	&	112	\\
00178750000	&	XRT/PC	&	2006-01-24 15:58:13	&	2006-01-24 15:59:01	&	15	&	200	\\
00178750001	&	XRT/PC	&	2006-01-24 19:05:00	&	2006-01-25 00:26:58	&	5501	&	11408	\\
00178750002	&	XRT/PC	&	2006-01-25 01:28:10	&	2006-01-25 22:55:56	&	22067	&	34399	\\
00178750003	&	XRT/PC	&	2006-01-26 00:05:00	&	2006-01-26 22:53:56	&	18569	&	115809	\\
00178750004	&	XRT/PC	&	2006-01-27 00:04:58	&	2006-01-27 19:37:58	&	14327	&	202206	\\
00178750005	&	XRT/PC	&	2006-01-28 00:08:40	&	2006-01-30 23:17:57	&	47631	&	288829	\\
00178750007     &       XRT/PC  &       2006-01-31 00:24:14     &       2006-02-01 23:24:57     &       44595   &       548963  \\
00178750009     &       XRT/PC  &       2006-02-02 00:41:55     &       2006-02-02 08:42:03     &       4759    &       722824  \\
00178750010	&	XRT/PC	&	2006-02-03 23:24:51	&	2006-02-03 23:35:58	&	667	&	891000	\\
00178750011	&	XRT/PC	&	2006-02-04 11:58:22	&	2006-02-04 20:28:57	&	4816	&	936210	\\
00178750012	&	XRT/PC	&	2006-02-05 13:40:08	&	2006-02-06 23:59:57	&	10779	&	1028716	\\
00178750013	&	XRT/PC	&	2006-02-07 01:11:20	&	2006-02-07 23:49:58	&	9483	&	1156588	\\
00178750014	&	XRT/PC	&	2006-02-08 01:12:24	&	2006-02-08 23:51:55	&	10310	&	1243053	\\
00178750015	&	XRT/PC	&	2006-02-09 06:07:56	&	2006-02-09 23:56:57	&	6443	&	1347185	\\
00178750016	&	XRT/PC	&	2006-02-10 00:04:58	&	2006-02-10 22:43:49	&	423	&	1411807	\\
00178750017	&	XRT/PC	&	2006-02-11 01:42:30	&	2006-02-13 23:03:56	&	7248	&	1504059	\\
00178750018	&	XRT/PC	&	2006-02-15 00:10:07	&	2006-02-15 22:55:56	&	11664	&	1844115	\\
00178750019	&	XRT/PC	&	2006-02-17 00:20:47	&	2006-02-17 23:07:57	&	15741	&	2017556	\\
  \hline
00178750000 & UVOT/V    & 2006-01-24 15:56:38 & 2006-01-24 15:59:14 &    153 &      106 \\
00178750000 & UVOT/V    & 2006-01-24 15:59:17 & 2006-01-24 16:11:34 &    725 &      265 \\
00178750001 & UVOT/B    & 2006-01-24 19:04:58 & 2006-01-24 19:07:20 &    139 &    11407 \\
00178750001 & UVOT/UVM2 & 2006-01-24 19:35:31 & 2006-01-24 19:39:15 &    221 &    13239$^{\mathrm{a}}$ \\
00178750001 & UVOT/UVM2 & 2006-01-24 19:39:18 & 2006-01-24 19:40:03 &     43 &    13466$^{\mathrm{a}}$ \\
00178750001 & UVOT/B    & 2006-01-24 20:45:53 & 2006-01-24 21:01:02 &    885 &    17461 \\
00178750001 & UVOT/UVW2 & 2006-01-24 21:01:08 & 2006-01-24 21:04:53 &    221 &    18376$^{\mathrm{a}}$ \\
00178750001 & UVOT/UVW2 & 2006-01-24 21:04:56 & 2006-01-24 21:08:41 &    221 &    18604$^{\mathrm{a}}$ \\
00178750001 & UVOT/UVW2 & 2006-01-24 21:08:44 & 2006-01-24 21:12:29 &    221 &    18832$^{\mathrm{a}}$ \\
00178750001 & UVOT/UVW2 & 2006-01-24 21:12:32 & 2006-01-24 21:16:17 &    221 &    19060$^{\mathrm{a}}$ \\
00178750001 & UVOT/B    & 2006-01-24 22:21:52 & 2006-01-24 22:37:01 &    885 &    23220 \\
00178750001 & UVOT/UVW2 & 2006-01-24 22:37:07 & 2006-01-24 22:40:52 &    221 &    24135$^{\mathrm{a}}$ \\
00178750001 & UVOT/UVW2 & 2006-01-24 22:40:55 & 2006-01-24 22:44:40 &    221 &    24363$^{\mathrm{a}}$ \\
00178750001 & UVOT/UVW2 & 2006-01-24 22:44:43 & 2006-01-24 22:48:28 &    221 &    24591$^{\mathrm{a}}$ \\
00178750001 & UVOT/UVW2 & 2006-01-24 22:48:31 & 2006-01-24 22:52:16 &    221 &    24819$^{\mathrm{a}}$ \\
00178750001 & UVOT/B    & 2006-01-24 23:55:52 & 2006-01-25 00:11:00 &    885 &    28860 \\
00178750001 & UVOT/UVW2 & 2006-01-25 00:11:07 & 2006-01-25 00:14:52 &    221 &    29775$^{\mathrm{a}}$ \\
00178750001 & UVOT/UVW2 & 2006-01-25 00:14:55 & 2006-01-25 00:18:40 &    221 &    30003$^{\mathrm{a}}$ \\
00178750001 & UVOT/UVW2 & 2006-01-25 00:18:43 & 2006-01-25 00:22:28 &    221 &    30231$^{\mathrm{a}}$ \\
00178750001 & UVOT/UVW2 & 2006-01-25 00:22:31 & 2006-01-25 00:26:16 &    221 &    30459$^{\mathrm{a}}$ \\
00178750002 & UVOT/V    & 2006-01-25 01:28:09 & 2006-01-25 01:55:00 &   1574 &    34398 \\
00178750002 & UVOT/V    & 2006-01-25 03:05:09 & 2006-01-25 03:32:00 &   1575 &    40217 \\
00178750002 & UVOT/V    & 2006-01-25 04:41:09 & 2006-01-25 05:08:00 &   1575 &    45977 \\
00178750002 & UVOT/V    & 2006-01-25 06:17:09 & 2006-01-25 09:57:00 &   4725 &    51737 \\
00178750002 & UVOT/V    & 2006-01-25 11:06:09 & 2006-01-25 14:46:00 &   4752 &    69078 \\
00178750002 & UVOT/V    & 2006-01-25 15:54:41 & 2006-01-25 22:56:00 &   7948 &    86389 \\
  \noalign{\smallskip}
  \hline
  \end{tabular}
  \end{center}
  \begin{list}{}{} 
  \item[$^{\mathrm{a}}$] Upper limit on detection. 
  \end{list} 
  \end{table*}

\appendix

	%%%%%%%%%%%%%%%%%%%%%%%%%%%%%%%%%%%%%%%%%%%%%%%%%%%%%%%%%%%%%%%%
	\section{Pile-up estimation in XRT/WT data \label{grb060124:pileup}} 
	%%%%%%%%%%%%%%%%%%%%%%%%%%%%%%%%%%%%%%%%%%%%%%%%%%%%%%%%%%%%%%%%

	\begin{figure} %%%%%%%%%%%%%%%%%%%%%%%%%%%%%%%%%%%%%%%%%%%%%%%%%%%%%%%%%%%%%%%%%
		\resizebox{\hsize}{!}{\includegraphics[angle=270]{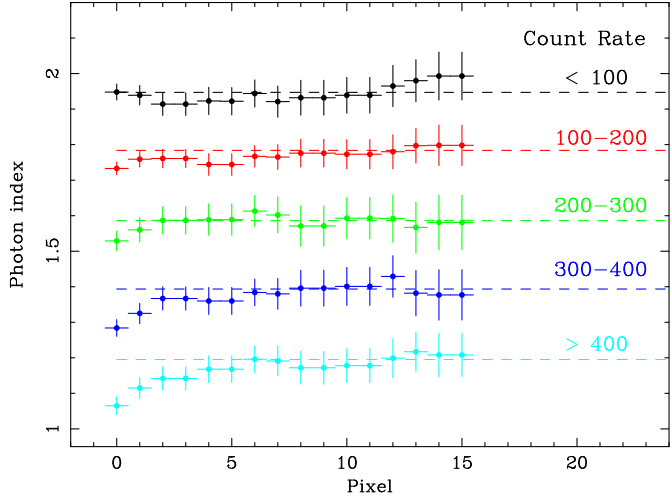}}
		\caption{Pile-up-driven spectral variations.
		XRT/WT photon index as a function of  observed source count rate and 
		size of the excluded region at the centre of the PSF (Sect.~\ref{grb060124:pileup}). 
		The photon indices are shifted along the ordinate direction for clarity. } 
 		\label{grb060124:xrtsppu}
	\end{figure}

The pile-up effect is a well known feature of CCD detectors \citep{Ballet99}.
It occurs when the input count rate is so high that
more than one X-ray photon is collected in a given $3\times3$ pixel 
region defining an
X-ray event in a single frame and the charges produced by the two separate
events are recorded as one. This effect causes groups of relatively soft
X-ray photons to be detected as one single high-energy photon.
The observed spectrum of an astrophysical source, therefore, is
distorted from its true form, with a general hardening of the slope
(decrease of the photon index).

Another feature of CCD detectors is that the charge produced by 
a cosmic ray is usually spread over many pixels, 
as opposed to charge produced by X-ray photons.
For this reason, a classification of the possible shapes of
the charge distribution is defined associating to each of
them a grade number: low grades correspond to confined
charge distributions and high grades to extended charge
distributions.
Cosmic rays are then recognized and rejected on the basis
of the event grade number.
Pile-up also occurs when two X-rays are collected in neighbouring
pixels in one frame. In this case groups of low grade events
can be recorded as a single high grade event.
Then, when pile-up occurs, the observed grade distribution
of the events will show a deficit of low grade events and
an excess of high grade events as compared with event grade
distribution of fainter sources. Since higher grade events
are rejected as cosmic rays, the pile-up effect causes
an apparent loss of flux, particularly from the centre of
the PSF where the count rate intensity is higher.

The standard way to correct for the pile-up effect is
to determine the extension of the PSF core affected by pile-up
and exclude from the analysis all the events that fell
within that region.
Based on the three observable effects we have described,
three different methods can be formulated to check for the presence
of pile-up and estimate the inner PSF region to exclude.
They involve
a study of the distortion of 1) the radial intensity profile; 
2) the spectrum;
3) the grade distribution.
The first method was discussed in detail
by \citet{vaughan2006:050315}, with particular attention to
Swift-XRT observations in PC mode, and it is widely applied
as standard correction technique for PC data.
The present analysis will focus on the following two
methods as favourite pile-up diagnostic tools
for Swift-XRT WT data.

Ground testing and calibration of the XRT integrated system 
carried out in 2002 Sep.\ 23--Oct.\ 4 at the Panter laboratory  
of the Max-Planck-Institut fur extraterrestrische Physik showed 
that WT mode data are affected by pile-up for point source  
input intensities above 250 counts s$^{-1}$.
To determine the extent to which pile-up distorts the spectral shape,
we selected 5 time intervals during which the observed source 
count rate was 
$<100$, 100--200, 200--300, 300--400, and $>400$ counts s$^{-1}$.
For each count rate range, spectra were extracted from
rectangular 40$\times$20-pixel regions
(40 pixels long along the image strip and 20 pixels wide)
with a region of increasing size (0$\times$20--15$\times$20 pixels)
excluded from its centre. 
The spectra were fit with an absorbed power law with
$N_{\rm H}$ fixed to the one obtained from the fit to the spectrum
extracted with the exclusion of 15 pixels (which is assumed to be 
unaffected by piled-up).
The spectra had 1433--2529 counts, depending on intensity cut and
exclusion region.
Figure~\ref{grb060124:xrtsppu} illustrates the pile-up-driven spectral
changes.
It demonstrates that in the range 0--100 counts s$^{-1}$ the
source is not piled-up, since no point deviates from the horizontal line,
which represents the best fit to the photon index values obtained in the
10--15 pixel exclusion range. For this intensity range, there is no need
to exclude any pixels from the extraction region.
The source is only affected by moderate pile-up for 100--300 counts s$^{-1}$; 
in this case it is necessary to exclude 1 pixel from 
the extraction region.
Finally, two and four pixels must be excluded for 300--400 counts s$^{-1}$) 
and $>400$ counts s$^{-1}$, respectively.

To determine the extent to which pile-up distorts the grades,
we studied the grade distribution for the same 5 time intervals
and the same exclusion regions as for the spectral study.
We derive analogous conclusions on the number of pixels to exclude from
the extraction regions as a function of observed source count rate.
The definitions of XRT grades for timing modes are different
from the ones for PC mode \citep{XRT}. Grades in WT span from 0 to 15, and
good events are likely associated to grades from 0 to 2 only.
If we assume that a good approximation of the expected grade 
distribution with no pile-up is obtained excluding 10 central pixels,  
the distributions we obtained show a deficit at grade zero (associated 
to an increase at higher grades) that becomes negligible if a 
region of at least 4-pixels is excluded.

\end{document}